\documentclass[prd,twocolumn, amsmath,amssymb,floatfix,nofootinbib]{revtex4}
\usepackage{mathrsfs,bm}
\usepackage{longtable,lscape}
\usepackage{txfonts}
\usepackage{indentfirst}
\usepackage{graphicx,color,dcolumn,booktabs}
\usepackage{multirow}
\usepackage{indentfirst}
\usepackage{multirow}
\usepackage{epsfig}
\usepackage{graphicx,color,dcolumn}
\usepackage{epstopdf}
\usepackage{amsmath}
\usepackage{diagbox}
\usepackage{changes}
\usepackage{threeparttable}
\usepackage{booktabs}
\usepackage{color}
\usepackage{amssymb}
\usepackage{cancel}
\definecolor{cover}{rgb}{0.77,0.87,0.88}
\definecolor{blueone}{rgb}{0.1,0.1,.7}
\definecolor{citec}{rgb}{0.14,0.47,0.09}
\definecolor{two}{rgb}{0.0,0.5,0.}
\definecolor{three}{rgb}{.5,.1,0.15}
\usepackage[bookmarks=true,bookmarksopen=false,plainpages=false,breaklinks=true,
   bookmarksnumbered=true,hypertexnames=false,
   filecolor=blue,urlcolor=three,menucolor=three,
   linkcolor=three,citecolor=blueone, colorlinks,
   anchorcolor=blue,runcolor=pink,frenchlinks=red
   pdfstartview=FitH,pdftitle=title,%
   pdfauthor=author]{hyperref}

\usepackage{relsize}
\usepackage{xspace}
\def\babar{\mbox{\slshape B\kern-0.1em{\smaller A}\kern-0.1em
    B\kern-0.1em{\smaller A\kern-0.2em R}}}

\newcolumntype{C}{>{$}c<{$}}
\AtBeginDocument{
\heavyrulewidth=.08em
\lightrulewidth=.05em
\cmidrulewidth=.03em
\belowrulesep=.65ex
\belowbottomsep=0pt
\aboverulesep=.4ex
\abovetopsep=0pt
\cmidrulesep=\doublerulesep
\cmidrulekern=.5em
\defaultaddspace=.5em
}

\allowdisplaybreaks[4]

\begin{document}
\title{Virtual state and two-pole structure: $\Xi(1690)$ and $\Xi(1820)$  from BESIII data on $\psi(3686)\to K^-\Lambda\bar{\Xi}^+$}
\author{Jun He}
\email{junhe@njnu.edu.cn}
\affiliation{School of Physics and Technology, Nanjing Normal University, Nanjing 210097, China}

\date{\today}
\begin{abstract}
We investigate the nature of the $\Xi(1690)$ and $\Xi(1820)$ resonances by analyzing the recent BESIII data on $\psi(3686) \to K^-\Lambda\bar{\Xi}^+$ within a coupled-channel rescattering framework. The calculation includes the channels $\bar{K}\Lambda$, $\bar{K}\Sigma^{(*)}$, $\pi\Xi^{(*)}$, $\eta\Xi^{(*)}$, and $K\Omega$. A direct fit to the data reveals that the $\Xi(1690)$ is best described as a virtual state located right below the $\bar{K}\Sigma$ threshold, a feature that naturally explains its very sharp peak observed in the $K^-\Lambda$ invariant mass spectrum. In contrast, the $\Xi(1820)$ exhibits a two-pole structure originating mainly from the coupled contributions of the $K\Sigma^*$ and $\pi\Xi^*$ channels, showing a two-pole pattern similar to that of the well-known $\Lambda(1405)$ and accounting for the unexpectedly broad width measured by BESIII. These findings provide new insights into the dynamical generation of hyperon resonances and highlight the importance of high-precision data from charmonium decays.
\end{abstract}

\maketitle

\section{ Introduction.} 
In the past two decades, exotic hadrons have become a central theme, driven by observations in heavy-flavor sectors~\cite{ParticleDataGroup:2024cfk}. The idea, however, emerged earlier in the strange sector, e.g., the $\Lambda(1405)$ as a $\bar{K}N$ bound state~\cite{Dalitz:1960du,Hyodo:2011ur,Oset:1997it,Hall:2014uca} and the rise and fall of the pentaquark $\Theta$~\cite{LEPS:2003wug}, which later inspired the discovery of hidden-charm pentaquarks~\cite{Wu:2010jy,Yang:2011wz,LHCb:2015yax,LHCb:2019kea}. Recently, renewed attention has focused on hyperon resonances, including the $\Sigma^*(1/2^-)$~\cite{Wu:2009tu,Wang:2024jyk,He:2025vij} and the $\Xi(1690)$ and $\Xi(1820)$, which are the subjects of this work. Early experiments established their existence and quantum numbers~\cite{Amsterdam-CERN-Nijmegen-Oxford:1976ezm,Teodoro:1978bu,Biagi:1986vs,BaBar:2008myc,Belle:2018lws}, but their nature remained unresolved. The new high-precision BESIII data on $\psi(3686)\to K^-\Lambda\bar{\Xi}^+$~\cite{BESIII:2015dvj,BESIII:2023mlv} reveal two prominent peaks corresponding to $\Xi(1690)$ and $\Xi(1820)$ standing out from the background, providing a clean environment to address this question.

Constituent quark models predict the masses of $\Xi(1690)$ and $\Xi(1820)$ around 1750~MeV~\cite{Capstick:1986ter,Glozman:1995fu}, significantly different from the PDG values~\cite{ParticleDataGroup:2024cfk}. Thus, they are natural candidates for dynamically generated states, as shown in the chiral unitary approach~\cite{Ramos:2002xh,Sarkar:2004jh}. A model with vector meson-baryon channels explains the narrow width of $\Xi(1690)$ by its small couplings to decay channels~\cite{Khemchandani:2016ftn}, and a pole near the $\bar{K}\Sigma$ threshold emerges as a hadronic molecule state~\cite{Sekihara:2015qqa}. A recent work following the Belle measurement reproduces $\Xi(1690)$ with a next-to-leading order chiral interaction~\cite{Feijoo:2023wua}. In the new BESIII measurement~\cite{BESIII:2023mlv}, the $\Xi(1690)$ peak lies just below the $\bar{K}\Sigma$ threshold and is very sharp, defying a standard Breit-Wigner description, a feature that has received little attention. This provides further evidence that $\Xi(1690)$ is a dynamically generated state rather than a conventional hyperon, a scenario we investigate here.

Likewise, the $\Xi(1820)$ also presents puzzles. The BESIII measurement shows a peak around $\Xi(1820)$ with an width of about 73~MeV~\cite{BESIII:2023mlv}, contrasting with the PDG average of $24\pm5$~MeV~\cite{ParticleDataGroup:2024cfk}. This large discrepancy has spurred theoretical work claiming the existence of two $\Xi(1820)$ states, as originally predicted within the chiral unitary approach~\cite{Sarkar:2004jh}. In that framework, the interaction of pseudoscalar mesons with $J^P=3/2^+$ decuplet baryons dynamically generates resonant states: a state around 1820~MeV and a wider pole in the 1800-1900~MeV region~\cite{Kolomeitsev:2003kt,Sarkar:2004jh}. Motivated by the BESIII measurement, Ref.~\cite{Molina:2023uko} revisited the $\Xi(1820)$ using coupled channels $\bar{K}\Sigma^*$, $\pi\Xi^*$, $\eta\Xi^*$, and $\Omega K$, obtaining a good description with two poles at $1824-31i$~MeV and $1875-130i$~MeV. Subsequent work~\cite{Liang:2024fsv} proposed $\Xi_c \to \pi^+(\pi^0,\eta)\pi\Xi^*$ reactions, where interference between the two resonances leads to a dip around 1850~MeV in the $\pi(\eta)\Xi^*$ invariant mass distributions. The two-pole structure is an intriguing phenomenon in hadron spectroscopy, first demonstrated for $\Lambda(1405)$ and later found in other cases such as $D_0^*(2300)$ and $K_1(1270)$~\cite{Jido:2003cb,Oller:2000fj,Albaladejo:2016lbb,Guo:2017jvc,Roca:2005nm,Geng:2006yb,He:2015cca}. These examples motivate our investigation into whether the $\Xi(1820)$ similarly arises from a two-pole scenario, which is the focus of this work.

Motivated by these considerations, we analyze the recent BESIII data on $\psi(3686) \to K^-\Lambda\bar{\Xi}^+$ to determine the nature of $\Xi(1690)$ and $\Xi(1820)$. Two features are key: (i) the $\Xi(1690)$ peak is very sharp and lies just below the $\bar{K}\Sigma$ threshold, defying a Breit-Wigner description and suggesting a dynamically generated state; (ii) the $\Xi(1820)$ exhibits an unexpectedly broad width, in sharp contrast to the PDG average, pointing to a possible two-pole structure. To address these issues, we employ a coupled-channel rescattering framework including $\bar{K}\Lambda$, $\bar{K}\Sigma^{(*)}$, $\pi\Xi^{(*)}$, $\eta\Xi^{(*)}$, and $K\Omega$, using the quasipotential Bethe-Salpeter equation (qBSE). Fitting the invariant mass spectra with the resulting amplitudes reveals the nature of these resonances. For reference, we also perform a conventional Breit-Wigner fit using effective Lagrangians.

\section{ Decay mechanism.} 
\begin{figure}[h!]
  \includegraphics[bb=80 642 540 765,clip,scale=0.52]{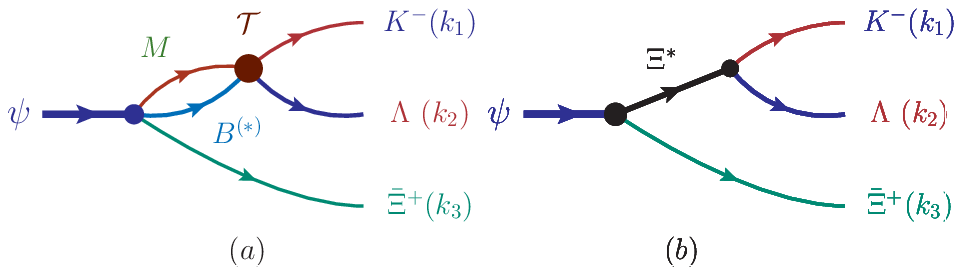}
  \caption{Feynman diagrams for $\psi(3686)\to K^-\Lambda\bar{\Xi}^+$. Panel (a) shows the rescattering mechanism, where the brown filled circles represent the coupled-channel amplitudes $\mathcal{T}$. Panel (b) depicts the conventional Breit-Wigner mechanism with intermediate resonances.}
  \label{diagram}
\end{figure}

The BESIII data show two peaks, $\Xi(1690)$ and $\Xi(1820)$, in the $K^-\Lambda$ invariant mass spectrum, while no structures are seen in the $K^-\bar{\Xi}^+$ or $\Lambda\bar{\Xi}^+$ spectra. Two possible decay mechanisms are illustrated in Fig.~\ref{diagram}.  

Our main focus is the rescattering scenario [Fig.~\ref{diagram}(a)], in which $\Xi(1690)$ and $\Xi(1820)$ arise as dynamically generated structures from coupled-channel interactions. Here, $\psi(3686)$ first decays into a meson-baryon pair, which then undergoes coupled-channel rescattering to produce the final $K^-\Lambda$ state. The included coupled channels are $\bar{K}\Lambda$, $\bar{K}\Sigma^{(*)}$, $\pi\Xi^{(*)}$, $\eta\Xi^{(*)}$, and $K\Omega$. For reference, we also consider the decay via intermediate Breit-Wigner resonances [Fig.~\ref{diagram}(b)]. In that case, $\psi(3686)$ decays into a $\Xi^*$ resonance accompanied by a $\bar{\Xi}^+$, followed by $\Xi^*\to K^-\Lambda$. This formalism, similar to that adopted in the experimental analysis, serves as a baseline for comparison.

The amplitudes of this Breit-Wigner formalism are derived from effective Lagrangians, with $\Xi(1690)$ and $\Xi(1820)$ treated as $1/2^-$ and $3/2^-$ resonances, respectively. The background is modeled as a wide Breit-Wigner resonance, either in the $K\Lambda$ channel with spin-parity $1/2^+$ or in the $\Lambda\bar{\Xi}^+$ channel with $0^-$. The explicit Lagrangians, coupling constants, and propagators are provided in the Sec.~S1 of Supplemental Material. Following the experimental analysis~\cite{BESIII:2023mlv}, form factors are included to account for off-shell effects of the $\Xi^*$ resonances. 

Before calculating the total decay amplitude in the rescattering scenario [Fig.~\ref{diagram}(a)], we first obtain the rescattering amplitudes for the coupled channels $\bar{K}\Lambda$, $\bar{K}\Sigma^{(*)}$, $\pi\Xi^{(*)}$, $\eta\Xi^{(*)}$, and $K\Omega$. For the interaction, we use the quasipotential Bethe-Salpeter equation (qBSE) combined with the one-boson-exchange model, where the potential is constructed from two vertices and a propagator. The vertices of pseudoscalar mesons ($\bar{K},\pi,\eta,K$) and exchanged vector mesons ($\rho,\omega,\phi,K^*$) are described by the Lagrangian
\begin{align}
{\cal L}_{VPP} = -ig \, {\rm Tr}\bigl\{ V_\mu [P, \partial^\mu P] \bigr\}, \label{Eq:VPP}
\end{align}
where the pseudoscalar and vector matrices are given by
\[
P = \begin{pmatrix}
\frac{1}{\sqrt{6}}\eta + \frac{1}{\sqrt{2}} \bm{\pi} \cdot \bm{\tau} & K \\
K^\dagger & -\frac{2}{\sqrt{6}}\eta
\end{pmatrix}, \quad
V = \begin{pmatrix}
\frac{1}{\sqrt{2}}\omega + \frac{1}{\sqrt{2}} \bm{\rho} \cdot \bm{\tau} & K^* \\
K^{*\dagger} & \phi
\end{pmatrix},
\]
with $\bm{\tau}$ the Pauli matrices acting in isospin space. This form of the Lagrangian has been derived in many approaches, such as hidden local symmetry~\cite{Bando:1984ej,Birse:1996hd} and the direct SU(3) symmetry construction with $\alpha = 1$~\cite{deSwart:1963pdg,Ronchen:2012eg}. The coupling constant is consistent across these formalisms and is given by $g = g_{\rho\pi\pi} / \sqrt{2}$, with $g_{\rho\pi\pi} \approx 6.1$~\cite{Bando:1984ej,Birse:1996hd,deSwart:1963pdg,Ronchen:2012eg,Janssen:1996kx,Oh:2004wp,Matsuyama:2006rp}.

The vertices of baryons with exchanged vector mesons are described by the following Lagrangians. For the interactions of octet baryons $B$ ($\Lambda$, $\Sigma$, $\Xi$) and decuplet baryons $B^*$ ($\Sigma^*$, $\Xi^*$, $\Omega$) with vector mesons $V_\nu$, we have
\begin{align}
{\cal L}_{BB V} &= g_{NN\rho} \, \bar{B} \gamma^\nu V_\nu B , \\
{\cal L}_{B^* B^* V} &= g_{\Delta\Delta\rho} \, \bar{B}^{*\mu} \gamma^\nu V_\nu B_\mu^* , \\
{\cal L}_{B^* B V} &= -i \frac{f_{\Delta N\rho}}{m_\rho} \, \bar{B}^{*\mu} \gamma^\nu \gamma^5 F_{\mu\nu} B ,
\end{align}
where $F_{\mu\nu} = \partial_\mu V_\nu - \partial_\nu V_\mu$ is the field strength tensor. Here we neglect the tensor coupling, as in Ref.~\cite{Lenske:2016ymj}; our calculation also confirms that its contribution is small, which facilitates a connection to the chiral unitary approach~\cite{Oset:1997it}. The coupling constants are taken as $g_{NN\rho}=g_{\rho\pi\pi}/2$~\cite{Janssen:1996kx,Matsuyama:2006rp,Ronchen:2012eg} and $g_{\Delta\Delta\rho}=g_{\rho\pi\pi}$ (close to the values in Ref.~\cite{Ronchen:2012eg}), while the transition coupling is fitted to $f_{\Delta N\rho}=-6.08$ from Ref.~\cite{Ronchen:2012eg}. The flavor structures of these vertices are provided in Sec.~S2 of the Supplemental Material.

Using the above Lagrangians, the vertices can be obtained, and the amplitudes for the potential kernel can be derived via standard Feynman rules. Following the method in Ref.~\cite{He:2019rva}, we directly implement the vertices $\Gamma$ and propagators $P$ into the code. The potential can be written using the Lagrangians and flavor wave functions as
\begin{equation}
{\cal V}_{{V}} = f_I \, \Gamma_{1\mu} \Gamma_{2\nu} P^{\mu\nu}_{{V}}, \label{V}
\end{equation}
where $f_I$ is the flavor factor, obtained from the flavor structures of the Lagrangians for mesons and baryons given above and the flavor wave functions (provided in  Sec.~S3 of the Supplemental Material). The propagator is defined as $P^{\mu\nu}_{V} = i{(-g^{\mu\nu} + q^\mu q^\nu / m_V^2)}/{(q^2 - m_V^2)}$ with $q = k - k'$ the four-momentum transfer and $m_V$ the vector meson mass. For the specific channels considered in this work, the interaction potentials take relatively simple forms. We present the explicit expressions as follows,
\begin{align}
{\cal V}_{MB\to MB}   &= C_{ij} g^2 (k+k')_\mu\, \bar{u}\,\gamma_\nu u\, i P^{\mu\nu}_V(q), \\
{\cal V}_{MB^*\to MB^*} &= C_{ij} g^2\, (k+k')_\mu\, \bar{u}^\rho\,\gamma_\nu u_\rho\, i P^{\mu\nu}_V(q), \\
{\cal V}_{MB\to MB^*} &= C_{ij} g^2\, (k+k')_\mu\, \bar{u}^\rho\,(\gamma_\nu q_\rho - g_{\rho\nu} q\mkern -9.5 mu / ) u\, i P^{\mu\nu}_V(q),
\end{align}
where $C_{ij}$ are numerical coefficients that depend on the specific coupled channels. Their explicit values for $MB\to MB$ and $MB^*\to MB^*$ are provided in Sec.~S3 of the Supplemental Material. In the above, $u$ ($\bar{u}$) denote the Dirac spinors of initial (final) octet baryons, while $u_\rho$ ($\bar{u}^\rho$) are the Rarita–Schwinger spinors for decuplet baryons. In the limit $q^2 \ll m_V^2$, neglecting the $q^\mu q^\nu$ term, the potential reduces to a contact interaction and coincides with that used in the chiral unitary approach~\cite{Ramos:2002xh,Sarkar:2004jh}.

The rescattering amplitude ${\cal T}$ [Fig.~\ref{diagram}(a)] is obtained by inserting the potential kernel ${\cal V}$ through the Bethe-Salpeter equation. Using the quasipotential approximation and partial-wave decomposition, the four-dimensional integral reduces to a one-dimensional integral equation for each $J^P$,
\begin{align}
&i{\cal T}^{J^P}_{\lambda_1,\lambda_2;\lambda'_1,\lambda'_2}({\rm p}',{\rm p}) \nonumber\\
&= i{\cal V}^{J^P}_{\lambda_1,\lambda_2;\lambda'_1,\lambda'_2}({\rm p}',{\rm p}) + \frac12 \sum_{\lambda''_1,\lambda''_2} \int \frac{{\rm p}''^2 d{\rm p}''}{(2\pi)^3} \nonumber\\
&\quad \cdot i{\cal V}^{J^P}_{\lambda_1,\lambda_2;\lambda''_1,\lambda''_2}({\rm p}',{\rm p}'') \, G_0({\rm p}'') \, i{\cal T}^{J^P}_{\lambda''_1,\lambda''_2;\lambda'_1,\lambda'_2}({\rm p}'',{\rm p}), \label{Eq: TJP}
\end{align}
The explicit can be found in our previous works~\cite{He:2014nya,He:2015mja,He:2017lhy,He:2015yva,He:2015cea}, and Sec.~S4 in the supplemental material. An exponential regulator $G_0({\rm p}'') \to G_0({\rm p}'') e^{-2(p''^2_l - m_l^2)^2/\Lambda^4}$ is introduced~\cite{He:2015mja}.

The decay amplitude is also partial-wave decomposed including parity~\cite{He:2025vij},
\begin{align}
&i{\cal M}_{\lambda_1,\lambda_2,\lambda_3;\lambda}(p_1,p_2,p_3)
\nonumber\\
&=\sum_{J^PM}\frac{2J+1}{4\pi}D^{J*}_{M\lambda_{21}}(\Omega_2)\int \frac{{\rm p}'^{2}d{\rm p}'}{(2\pi)^3} \nonumber\\
&\cdot\sum_{\lambda'_1\lambda'_2} i{\cal T}^{J^P}_{\lambda_1,\lambda_2;\lambda'_1\lambda'_2}({\rm p}'_j,s_{12})  \ G_0({\rm p}') i{\cal A}^{J^PM}_{\lambda'_1\lambda'_2;\lambda_3;\lambda}({\rm p}'_j,\Omega_3,s_{12}),
\end{align}
where $\Omega_2,\Omega_3$ are solid angles of $\Lambda,\bar{\Xi}^+$, $s_{12}$ is the $K^-\Lambda$ invariant mass, and $D^{J*}_{M\lambda_{21}}$ is a Wigner $D$-function.
The amplitude ${\cal A}$ is described by the effective Lagrangians
\begin{align}
{\cal L}_{\psi\to MB^* \bar{\Xi}^+} &= g_{B^*M} \; \bar{B}_\mu M \gamma_\nu \psi^\mu \partial^\nu \bar{\Xi}^+, \\
{\cal L}_{\psi\to MB \bar{\Xi}^+} &= g_{BM} \; \bar{B} M \psi_\mu \partial^\mu \bar{\Xi}^+,
\end{align}
It should be noted that we consider the decay $\psi(3686)\to K^-\Lambda\bar{\Xi}^+$, and the rescattering occurs in the $K^-\Lambda$ channel, which has total isospin $I=1/2$. In principle, the intermediate states should include all coupled channels: $MB^{(*)}=\bar{K}\Lambda$, $\bar{K}\Sigma^{(*)}$, $\pi\Xi^{(*)}$, $\eta\Xi^{(*)}$, and $K\Omega$, with isospin $I=1/2$. As shown in Ref.~\cite{Duan:2024ygq}, $\psi(3686)$, being a $c\bar{c}$ state, is an SU(3) singlet (for $u,d,s$ quarks). Consequently, the decay amplitudes to the $MB$ channels are related within that sector, and likewise for the $MB^*$ channels, but there is no relation between the two sectors. The fit results show that the contributions from the $MB$ channels are numerically very small, while the $MB^*$ channels play an essential role in generating the near-threshold structures observed in the data. We therefore retain only the $MB^*$ channels as intermediate channels. Their relative contributions for $I=1/2$ states, based on SU(3) symmetry, are $\bar{K}\Sigma^* : \pi\Xi^* : \eta\Xi^* : K\Omega = 1 : -1 : -1 : \sqrt{2}$~\cite{deSwart:1963pdg,Duan:2024ygq}.
For the rescattering mechanism, the form factors adopted in the BESIII analysis~\cite{BESIII:2023mlv} centered at the resonance pole masses $m_{\Xi^*}$ are not suitable.  Therefore, we introduce a different form factor at the decay vertices, based on the $K\Lambda$ threshold: $F(s_{12}) = \Lambda^4/(\Lambda^4 + (s_{12} - m_{\rm thr}^2)^2)$ where $m_{\rm thr}=m_K+m_\Lambda$ is the $K\Lambda$ threshold mass, and cutoff $\Lambda = 1$~GeV.

The partial-wave vertex ${\cal A}^{J^PM}$ is obtained from the amplitudes ${\cal A}$ via a partial-wave decomposition:
\begin{align}
{\cal A}^{J^PM}_{\lambda'_1,\lambda'_2;\lambda_3;\lambda}
&= \int d\Omega'_2 \Big[ D^{J}_{M,\lambda'_{21}}(\Omega'_2) \, {\cal A}_{\lambda'_1,\lambda'_2;\lambda_3;\lambda}
\nonumber\\
&\qquad + \eta' D^{J}_{M,-\lambda'_{21}}(\Omega'_2) \, {\cal A}_{-\lambda'_1,-\lambda'_2;\lambda_3;\lambda}(\Omega'_2) \Big],
\end{align}
where $\lambda'_{21} = \lambda'_2 - \lambda'_1$ is the helicity difference of the two intermediate particles, and $\eta' = P P_1 P_2 (-1)^{J-J_1-J_2}$ is the parity factor, with $P$ the intrinsic parity of the system, $P_1,P_2$ the parities of the two constituents, and $J,J_1,J_2$ the total and individual spins.

The experimental analysis suggests that the data can be described by two resonances plus a background contribution. Following the experimental procedure, the total decay amplitude is expressed as
\begin{align}
{\cal M} = e^{i\pi\phi_{\Xi(1690)}} {\cal M}_{\Xi(1690)} + e^{i\pi\phi_{\Xi(1820)}} {\cal M}_{\Xi(1820)} + {\cal M}_{\text{bk}},
\end{align}
where $e^{i\pi\phi_{\Xi^*}}$ and ${\cal M}_{\Xi^*}$ denote the relative phase and amplitude associated with the resonance $\Xi^*$, each of which will be described either by Breit-Wigner intermediate states or by rescattering effects. The background contribution is modeled as a wide Breit-Wigner resonance in two scenarios, following the experimental analysis~\cite{BESIII:2023mlv}.

With the decay amplitude, the differential decay width is given by
\begin{align}
d\Gamma = \frac{1}{2M_\psi} \sum |{\cal M}|^2 d\Phi,
\end{align}
where $M_\psi$ is the mass of the initial $\psi(3686)$ meson, and $d\Phi$ denotes the Lorentz-invariant phase space element. In this work, the phase space element $d\Phi$ is simulated using the method proposed in Ref.~\cite{James:1968gu}, implemented in the Julia programming language. The corresponding Julia package \texttt{DalitzPlot.jl}, which simulates decays based on our qBSE approach, is available on GitHub~\cite{code}. A total of $10^7$ events are generated. The momenta of the final-state particles are sampled via a Monte Carlo approach that strictly enforces energy-momentum conservation. From these simulated momenta, the invariant mass spectra are constructed and subsequently used in the analysis.

\section{ Numerical Results.}\label{Sec:results} 
We first fit the BESIII invariant mass spectra of $K^-\Lambda$, $K^-\bar{\Xi}^+$, and $\Lambda\bar{\Xi}^+$ using Breit-Wigner resonances $\Xi(1690)$ and $\Xi(1820)$, with the background as a wide Breit-Wigner resonance in $K^-\Lambda$ or $\Lambda\bar{\Xi}^+$ (second and third columns of Table~\ref{tab:fit}).
\renewcommand\tabcolsep{0.19cm}
\renewcommand{\arraystretch}{1.5}
\begin{table}[h!]
\caption{Fit parameters and $\chi^2$ values for different scenarios: BW (Breit-Wigner) and RS (rescattering). Backgrounds (bk) are added in either $K^-\Lambda$ or $\Lambda\bar{\Xi}^+$ spectra, as indicated. Strength parameters $g$ are in relative units (only magnitudes meaningful), phases $\phi$ dimensionless, masses $m$ and widths $\Gamma$ in MeV. Fit quality: $\chi^2_{\rm tot}/{\rm ndf}$ (total chi-squared per degree of freedom).\label{tab:fit}}
\begin{tabular}{ccccc}
		\toprule
Fits                     &BW($K^-\Lambda$)    &BW($\Lambda\bar{\Xi}^+$) & RS($K^-\Lambda$)    &RS($\Lambda\bar{\Xi}^+$) \\\hline
$g_{\Xi(1690)}$              &$0.247_{-0.004}^{+0.004}$   &$0.194_{-0.003}^{+0.004}$        &$35.33_{-0.33}^{+0.46}$      &$32.51_{-0.75}^{+0.26}$   \\
$\phi_{\Xi(1690)}$          &$1.948_{-0.003}^{+0.004}$   &$0.282_{-0.003}^{+0.002}$        &$0.338_{-0.037}^{+0.199}$      &$0.822_{-0.056}^{+0.134}$     \\
$m_{\Xi(1690)}$             &$1682.9_{-4.8}^{+3.0}$   &$1667.3_{-5.5}^{+2.1}$        &$--$          &$--$       \\
$\Gamma_{\Xi(1690)}$    &$134.8_{-2.8}^{+3.2}$   &$92.0_{-2.3}^{+2.0}$        &$--$          &$--$         \\
$g_{\Xi(1820)}$               &$4.834_{-0.056}^{+0.093}$   &$4.686_{-0.070}^{+0.088}$        &$65.16_{-1.20}^{+1.60}$      &$62.94_{-2.98}^{+0.67}$       \\
$\phi_{\Xi(1820)}$           &$1.165_{-0.025}^{+0.031}$   &$1.010_{-0.001}^{+0.031}$        &$1.062_{-0.323}^{+0.238}$      &$0.565_{-0.191}^{+0.028}$       \\
$m_{\Xi(1820)}$               &$1820.8_{-1.9}^{+1.9}$   &$1821.2_{-2.6}^{+2.0}$        &$--$          &$--$    \\
$\Gamma_{\Xi(1820)}$     &$94.0_{-2.2}^{+2.0}$   &$100.0_{-2.2}^{+2.8}$        &$--$          &$--$     \\
$g_{\rm bk}$                  &$1.054_{-0.020}^{+0.023}$   &$9.128_{-0.122}^{+0.212}$        &$0.603_{-0.034}^{+0.039}$      &$11.39_{-0.54}^{+0.16}$       \\
$m_{\rm bk}$                &$2034.9_{-8.9}^{+9.4}$   &$2846.0_{-5.6}^{+6.3}$        &$1911.4_{-32.7}^{+32.6}$      &$2885.4_{-6.0}^{+5.6}$          \\
$\Gamma_{\rm bk}$           &$1489.3_{-32.8}^{+29.0}$   &$860.1_{-19.8}^{+12.5}$        &$1394.4_{-82.4}^{+89.5}$      &$1106.1_{-16.9}^{+56.1}$            \\
${\chi^2_{\rm tot}}/{\rm ndf}$    &$1.7059$   &$2.0103$        &$2.4127$      &$1.6671$       \\
		\bottomrule
	\end{tabular}
\end{table}
This fit has 11 parameters (strength, phase, mass, width for each $\Xi^*$, plus three background parameters). The $\chi^2 = \sum_{i=1}^{N_{\rm bin}} (n_i - v_i)^2 / v_i$ follows the experimental definition~\cite{BESIII:2023mlv}. The strength parameters, $g_{\Xi(1690)}$, $g_{\Xi(1820)}$, and $g_{\rm bk}$, absorb coupling constants and are scaled to the overall yield; only their relative values matter. We use the Julia package \texttt{optimize.jl} for the fit. With a background in $K^-\Lambda$, we obtain $\chi^2_{\rm tot}/\text{ndf}=1.7059$, close to the experimental result, while a background  in $\Lambda\bar{\Xi}^+$ gives $\chi^2_{\rm tot}/\text{ndf}=2.0103$. Our main focus, however, is the rescattering scenario ( fourth and fifth columns of Table~\ref{tab:fit}), where the $\Xi(1690)$ and $\Xi(1820)$ are dynamically generated; only strength and phase parameters are fitted there.Remarkably, with the background in $\Lambda\bar{\Xi}^+$, the rescattering scenario yields $\chi^2_{\rm tot}/\text{ndf}=1.6671$, an excellent fit that demonstrates its ability to describe the data.

The results for the Breit-Wigner scenario are illustrated in Fig.~\ref{fig:BW} (the full set of results for the three invariant mass spectra of $K^-\Lambda$, $K^-\bar{\Xi}^+$, and $\Lambda\bar{\Xi}^+$ is provided in Sec.~S5 of the Supplemental Material). Our treatment is similar to that adopted in the experimental analysis, but with effective Lagrangians. The fit includes three components: two Breit-Wigner resonances for $\Xi(1690)$ and $\Xi(1820)$, and one for the background.

\begin{figure}[h!]
\centering
\includegraphics[bb=0 0 400 700,clip,scale=0.55]{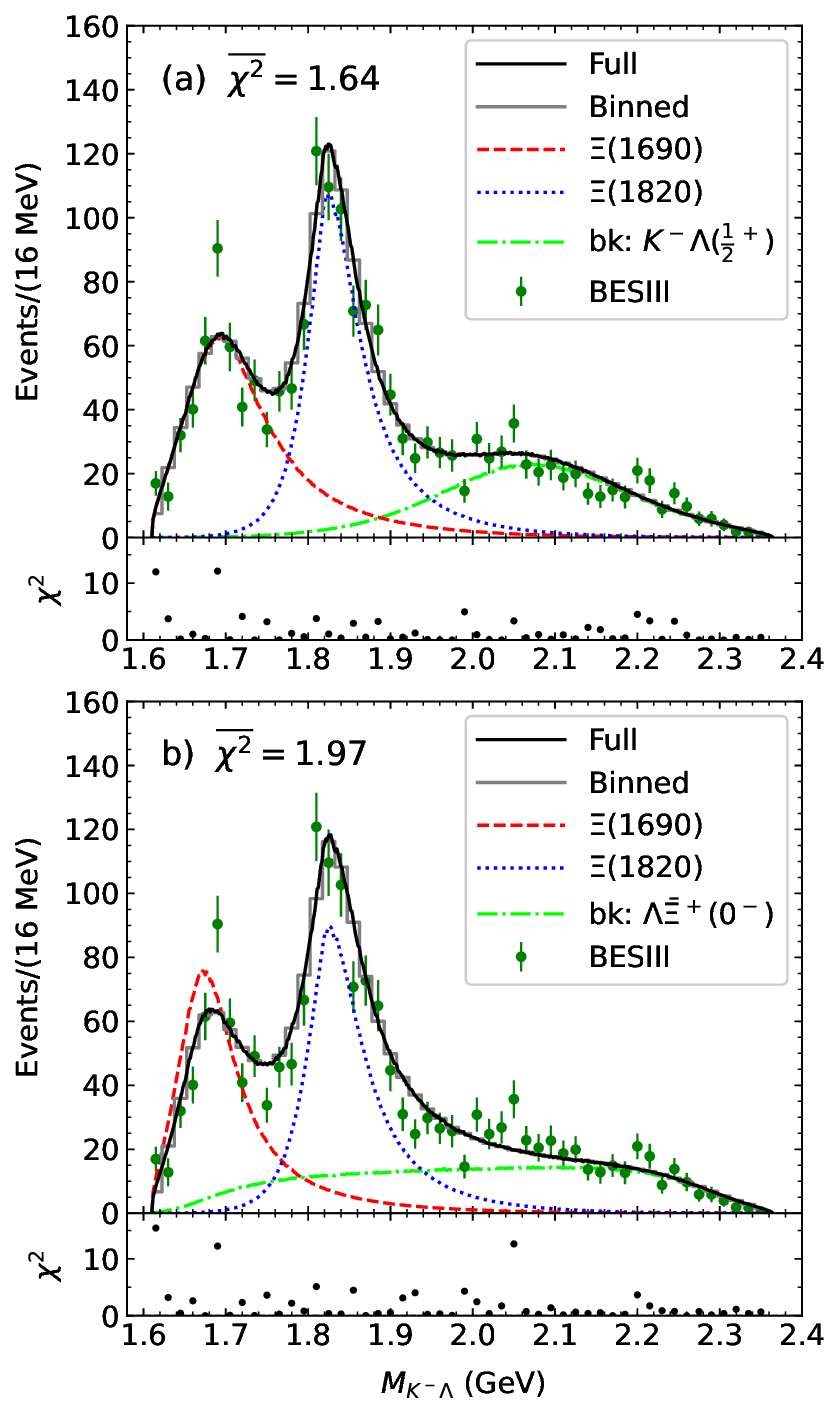}
\caption{$K^-\Lambda$ invariant mass spectra in the Breit-Wigner scenario. The curves correspond to the full model (solid black), $\Xi(1690)$ (red dashed), $\Xi(1820)$ (blue dotted), and background (green dash-dotted). The binned results are also shown as a gray histogram. The data are taken from the BESIII Collaboration~\cite{BESIII:2023mlv}. The scatter points in the lower part of each panel represent the individual $\chi^2$ contributions of the corresponding data points. Panel (a) shows the fit with a background in $K^-\Lambda$ ($1/2^+$), and panel (b) shows the fit with a background in $\Lambda\bar{\Xi}^+$ ($0^-$).}
\label{fig:BW}
\end{figure}

As in the experimental analysis, the $K^-\Lambda$ invariant mass spectrum can be well reproduced with two $\Xi^*$ states and a background in $K^-\Lambda$ with $1/2^+$. The average $\chi^2$ values for the $K^-\Lambda$, $K^-\bar{\Xi}^+$, and $\Lambda\bar{\Xi}^+$ spectra obtained from this fit are 1.64, 1.12, and 2.06, respectively (see Fig.~\ref{fig:BW}(a) for $K^-\Lambda$; the $K^-\bar{\Xi}^+$ and $\Lambda\bar{\Xi}^+$ spectra are shown in the Supplement Material). These values are close to those reported by the BESIII collaboration (1.46, 1.10, and 2.16)~\cite{BESIII:2023mlv}. As shown in Fig.~\ref{fig:BW}(a), the peaks around 1700 and 1800~MeV in the $K^-\Lambda$ spectrum clearly originate from the two Breit-Wigner resonances, while the background mainly contributes to the higher energy region. Replacing the background in $K^-\Lambda$ with one in $\Lambda\bar{\Xi}^+$ ($0^-$) yields a slightly improved $\chi^2$ for the $\Lambda\bar{\Xi}^+$ spectrum (from 2.06 to 1.89) but worsens the fits for $K^-\Lambda$ and $K^-\bar{\Xi}^+$ ($\chi^2$ values increase to 1.97 and 1.72, respectively), consistent with the experimental finding that the background in $K^-\Lambda$ with $1/2^+$ is the best choice~\cite{BESIII:2023mlv}.

An important observation is that the experimental peak around 1700~MeV is very sharp. Neither of the fits shown in Fig.~\ref{fig:BW}(a) and (b) can reproduce such a sharp peak; in fact, even the original experimental analysis using Breit-Wigner resonances failed to describe it~\cite{BESIII:2023mlv}. This provides strong evidence that the peak does not originate from a conventional Breit-Wigner resonance, but rather points to a non-resonant or dynamically generated nature.

\begin{figure*}[t]
\centering
\includegraphics[bb=5 0 1050 700,clip,scale=0.48]{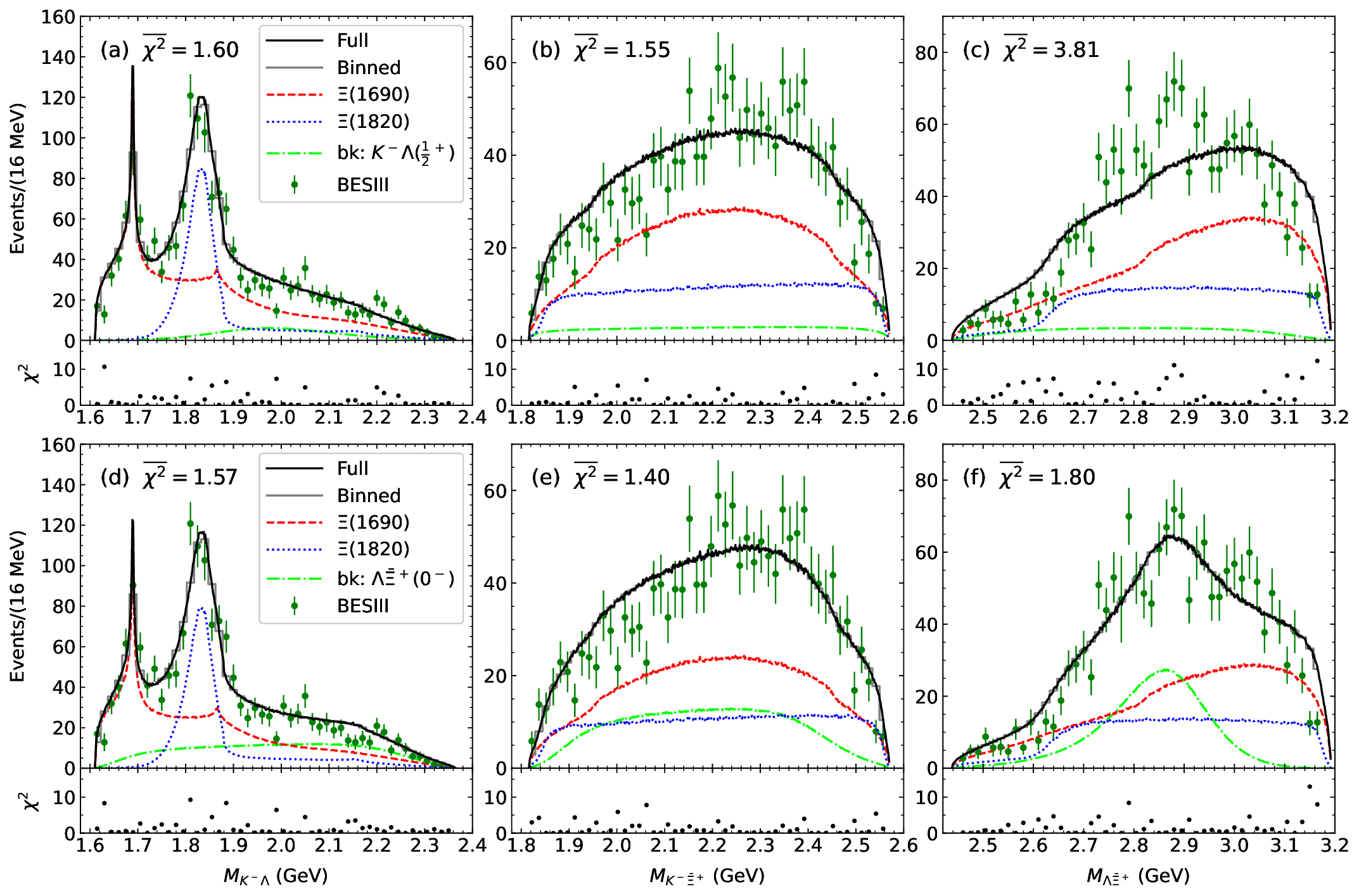}
\caption{Invariant mass spectra in the rescattering scenario. Panels (a, b, c) corresponds to the fit with a background in $K^-\Lambda$ ($1/2^+$), and panels (d, e, f) to the fit with a background in $\Lambda\bar{\Xi}^+$ ($0^-$). Other notations are the same as in Fig.~\ref{fig:BW}.}
\label{fig:RS}
\end{figure*}

Now we replace the Breit-Wigner resonances with the rescattering mechanism. Specifically, we consider coupled-channel interactions involving $\bar{K}\Lambda$, $\bar{K}\Sigma^{(*)}$, $\pi\Xi^{(*)}$, $\eta\Xi^{(*)}$, and $K\Omega$. The lower peak, $\Xi(1690)$, originates mainly from the interaction of octet baryons with pseudoscalar mesons, i.e., $\bar{K}\Lambda$, $\bar{K}\Sigma$, $\pi\Xi$, and $\eta\Xi$, in an $S$-wave, which forms a system with $J^P=1/2^-$, consistent with the experimental suggestion. The higher peak, $\Xi(1820)$, arises primarily from the channels $\bar{K}\Sigma^*$, $\pi\Xi^*$, and $\eta\Xi^*$ in an $S$-wave, which forms a system with $J^P=3/2^-$, also consistent with the experimental suggestion. In the calculation, the couplings among all these channels are fully considered. We find that with a cutoff $\Lambda = 1.15$~GeV, a peak near 1800~MeV is generated, as shown in Figs.~\ref{fig:RS}(a) and (d).

Unlike the $\Xi(1820)$, which can be reproduced by both a Breit-Wigner resonance and the rescattering mechanism, the $\Xi(1690)$ cannot be well fitted by a Breit-Wigner resonance, as shown in Fig.~\ref{fig:BW} and in the original experimental analysis. The main issue is that the experimental peak is too sharp and does not follow a standard Breit-Wigner shape. Such a sharp peak is a natural feature of a molecular state near threshold, and can be well described by our rescattering mechanism. While a Flatt\'e-like parametrization can also produce a similar sharp peak near threshold, it remains purely phenomenological and does not provide access to the underlying dynamics, which are naturally obtained in our qBSE approach. To find the best description of the data, we adjust the cutoff for the $\bar{K}\Sigma$ channel and set it to $0.5$~GeV. As shown in Figs.~\ref{fig:RS}(a) and (d), a very sharp peak is reproduced from the coupled-channel rescattering. The maximum value of the peak is slightly higher than the experimental data points, which is due to the binning of the data at intervals of 16~MeV. After binning the theoretical results (gray histogram in Figs.~\ref{fig:RS}), the binned results are found to fit the data well.

The rescattering amplitudes with spin parity $1/2^-$ and $3/2^-$ can reproduce the two peaks around 1700 and 1800~MeV. Unlike the fit with Breit-Wigner resonances, the $1/2^-$ contribution appears in almost the entire energy region of the $K^-\Lambda$ invariant mass spectrum. Besides providing the $\Xi(1690)$ peak, a clear effect of this contribution is also seen in the $\Xi(1820)$ peak, and it becomes even more important in the other two invariant mass spectra. If we choose the background in the $K^-\Lambda$ channel, the background contribution is suppressed due to the $1/2^-$ rescattering contribution. However, with such a small background, the $\Lambda\bar{\Xi}^+$ invariant mass spectrum cannot be well reproduced, yielding an average $\chi^2$ of about 3.81. To improve the fit in all three invariant mass spectra, we move the background to the $\Lambda\bar{\Xi}^+$ channel. An excellent fit is then achieved with two rescattering contributions ($1/2^-$ and $3/2^-$) and a $\Lambda\bar{\Xi}^+(0^-)$ background. The average $\chi^2$ values for $K^-\Lambda$, $K^-\bar{\Xi}^+$, and $\Lambda\bar{\Xi}^+$ become 1.57, 1.40, and 1.80, respectively, with a total reduced $\chi^2$ of 1.667, which is very close to the experimental values.

\begin{figure}[htbp]
\centering
\includegraphics[bb=5 0 1200 400,clip,scale=0.515]{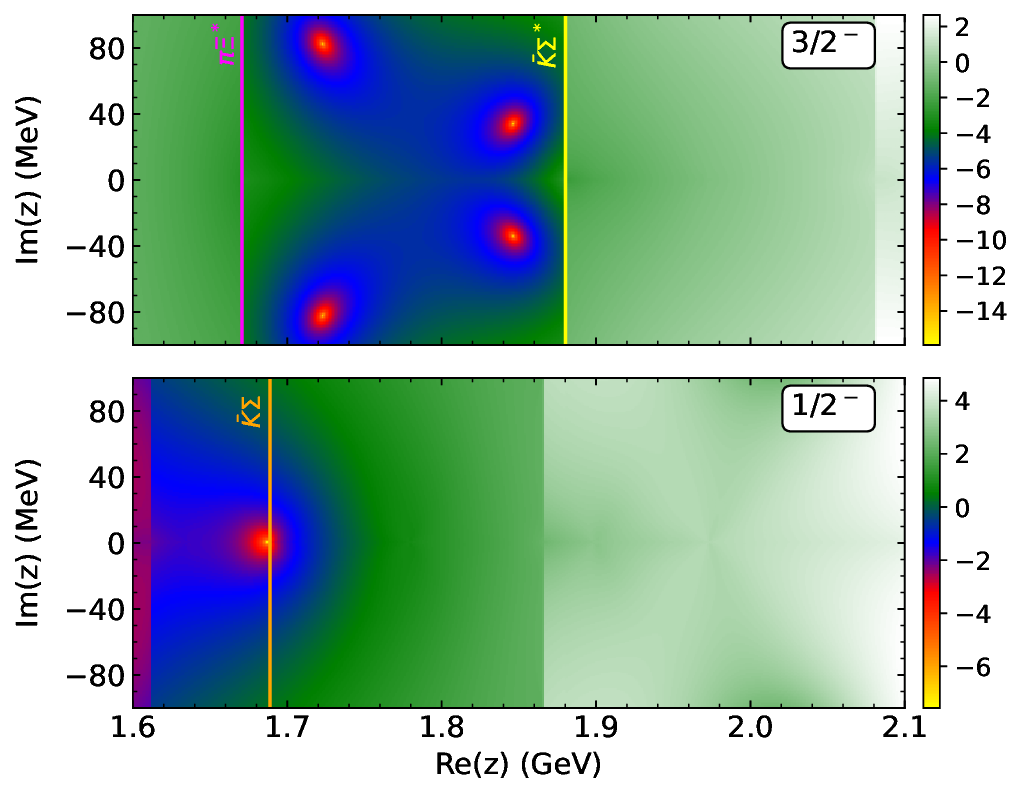}
\caption{Poles with spin-parities $3/2^-$ (upper panel) and $1/2^-$ (lower panel) from the coupled-channel interactions. The colorbox represents $\log|1 - V(z) G(z)|$. Vertical dashed lines mark the positions of the $\bar{K}\Sigma^*$, $\pi\Xi^*$ (upper panel) and $\bar{K}\Sigma$ (lower panel) thresholds.}
\label{fig:pole}
\end{figure}

In the above, with rescatterings in the $K^-\Lambda$ channel, the experimental data can be well reproduced. The coupled-channel interactions including $\bar{K}\Lambda$, $\bar{K}\Sigma^{(*)}$, $\pi\Xi^{(*)}$, $\eta\Xi^{(*)}$, and $K\Omega$ reproduce the correct invariant mass spectra. Now, we provide the poles of these coupled-channel interactions in the complex energy plane, as shown in Fig.~\ref{fig:pole}, to discuss the nature of the $\Xi(1690)$ and $\Xi(1820)$.

To locate the poles, we adopt the method of Ref.~\cite{Roca:2005nm}. For each two-body channel, we choose the first (physical) Riemann sheet below its threshold and the second (unphysical) sheet above threshold; this convention ensures that the resulting pole positions and half-widths can be meaningfully compared to Breit-Wigner parameters extracted from the real axis. In the $3/2^-$ partial wave (upper panel of Fig.~\ref{fig:pole}), two poles are clearly visible. The pole near $1846-34i$~MeV lies close to the $K\Sigma^*$ threshold (approximately $1880$~MeV). A single-channel analysis reveals that this pole originates from a bound state of the $K\Sigma^*$ interaction; when coupled channels are switched on, the bound state acquires an imaginary part because it can decay into other channels (e.g., $\pi\Xi^*$), thereby turning into a resonance with a finite width. The second pole at $1724-80i$~MeV is located above the $\pi\Xi^*$ threshold. Since it is a resonance mainly generated by the $\pi\Xi^*$ interaction, it carries an imaginary part already in the single-channel calculation, and this width is further enlarged by coupled-channel effects; hence, it naturally has a larger width. One can also note that the real part of the higher pole ($1846$~MeV) is higher than the Breit-Wigner mass and those reported in Ref.~\cite{Liang:2024fsv}, and the existence of the lower pole shifts the peak slightly lower relative to the experimental one. The co-existence of these two poles in the same $J^P$ partial wave constitutes a two-pole structure, which underlies the observed line shape in the $K^-\Lambda$ invariant mass spectrum: the higher pole dominates the peak near $1820$~MeV, while the lower pole contributes with a broader width.

In the $1/2^-$ partial wave (lower panel), only one pole is found near $1690$~MeV, which is essential for reproducing the very sharp peak in the data. This pole lies almost exactly at the $K\Sigma$ threshold. A detailed analysis of its location in the complex plane reveals that it is not a state on the physical sheet but a virtual state: it resides on the second Riemann sheet of the $K\Sigma$ channel at $1687.1$~MeV, approximately $1.7$~MeV below the $K\Sigma$ threshold. Virtual states are characterized by a pole on the real axis just below threshold on the unphysical sheet, producing a cusp-like enhancement in the scattering amplitude rather than a typical Breit-Wigner peak. To visualize this pole, in the lower panel we specifically choose the second Riemann sheet for the $K\Sigma$ channel, while for the other channels we adopt the same prescription as used in the $3/2^-$ case. A clear virtual state then appears as a localized structure below the $K\Sigma$ threshold. This virtual state is a direct consequence of the attractive $\bar{K}\Sigma$ interaction in the $S$-wave, which is too weak to form a bound state but sufficiently strong to produce a near-threshold singularity.  This virtual state is robust against variations of the cutoff parameter, as discussed in Sec.~S6 of the supplemental material. This stability, together with the fact that the data strongly favor this scenario over bound-state interpretation, supports the virtual-state nature of the $\Xi(1690)$.

\section{ Summary} 
The Breit-Wigner fits, similar to the experimental analysis, reproduce the overall spectra reasonably well but fail to describe the very sharp $\Xi(1690)$ peak near the $\bar{K}\Sigma$ threshold. In contrast, the rescattering mechanism naturally produces this sharp peak, which arises from a virtual state located about $1.7$~MeV below the threshold in the $1/2^-$ partial wave. The current BESIII data thus support the interpretation that $\Xi(1690)$ is not a conventional three‑quark state but a dynamically generated virtual state from $\bar{K}\Sigma$ interactions. Future high‑precision data will be essential to further test this scenario.

For $\Xi(1820)$, the rescattering calculation yields a two-pole structure in the $3/2^-$ partial wave: a pole at $1846-34i$~MeV below the $K\Sigma^*$ threshold (originating from a bound state) and another at $1724-80i$~MeV above the $\pi\Xi^*$ threshold (originating from a resonance). This explains the broad BESIII width, contrasting with the narrower PDG average. The pattern is consistent with the well-known $\Lambda(1405)$ two-pole scenario, where a higher pole lies close to the real axis and a lower, broader pole appears~\cite{Oller:2000fj,Jido:2003cb,He:2015cca}. We note that previous chiral unitary approach calculations (e.g., Ref.~\cite{Molina:2023uko}), which do not perform a combined fit to the three invariant mass spectra, give a somewhat different two-pole pattern ($1824-31i$~MeV and $1875-130i$~MeV). This difference highlights the sensitivity of the pole pattern to the underlying dynamics and the importance of using high-precision data to constrain theoretical models. Although our dynamical generation picture provides a natural explanation for the two-pole pattern, we do not exclude the possibility that the $\Xi(1820)$ may have a sizeable conventional three-quark component. A quantitative distinction between dynamically generated and elementary baryon states can be investigated through various theoretical approaches, such as the Weinberg compositeness criterion. Further experimental and theoretical studies are needed to discriminate between these scenarios.

In summary, our analysis favors that $\Xi(1690)$ is a virtual state from $\bar{K}\Sigma$ interactions, and $\Xi(1820)$ exhibits a $\Lambda(1405)$-like two-pole structure, reinforcing the dynamical generation of strange baryon resonances.

\vskip 10pt \noindent {\bf Acknowledgement} This project is supported by the National Natural Science Foundation of China (Grant No. 12475080)

\newpage
\addtocontents{toc}{\string\tocdepth@munge}
\section{S1. Lagrangians for Breit-Wigner scenario}

The amplitudes for Fig.~1(b) are derived from effective Lagrangians, with $\Xi(1690)$ and $\Xi(1820)$ treated as $1/2^-$ and $3/2^-$ resonances, respectively. The background is modeled as a wide Breit-Wigner resonance: a $b(1/2^+)$ wave in $K\Lambda$ or a $b(0^-)$ wave in $\Lambda\bar{\Xi}^+$.

The Lagrangians for the decays of $\psi(3686)$ to intermediate Breit-Wigner resonances are written as
\begin{align}
{\cal L}_{\psi\to\Xi^*(3/2^-)\bar{\Xi}^+} &= g_{3/2^-}\, \bar{\Xi}^{*-}_\mu \gamma_\nu \psi^\mu \partial^\nu \bar{\Xi}^+ ,\\
{\cal L}_{\psi\to\Xi^*(1/2^-)\bar{\Xi}^+} &= g_{1/2^-}\, \bar{\Xi}^{*-} i\gamma_5 \psi_\mu \partial^\mu \bar{\Xi}^+ ,\\
{\cal L}_{\psi\to b(1/2^+)\bar{\Xi}^+} &= g_{1/2^+}\, \bar{b} \gamma^\mu \psi_\mu \bar{\Xi}^+ ,\\
{\cal L}_{\psi\to b(0^-)\bar{\Xi}^+} &= g_{0^-}\, \bar{b} \psi_\mu \partial^\mu K^- ,
\end{align}
where $\Xi^{*-}_\mu$, $\Xi^-$, $\bar{\Xi}^+$, $\psi$ and $K^-$ denote the fields of $\Xi^{*-}$, $\Xi^-$, $\bar{\Xi}^+$, $\psi(3686)$ and $K^-$, respectively, and the $g_{J^P}$'s are the corresponding coupling constants. In the last two Lagrangians, $b$ represents a generic intermediate resonance as background in the $K\Lambda$ or $\Lambda\bar{\Xi}^+$ channel.

The Lagrangians for the subsequent decays of the intermediate resonances are written as
\begin{align}
{\cal L}_{\Xi^*(3/2^-)\to\Lambda K^-} &= g'_{3/2^-}\, \bar{\Lambda} \gamma_5 \partial^\mu K^- \Xi^{*-}_\mu , \\
{\cal L}_{\Xi^*(1/2^-)\to\Lambda K^-} &= g'_{1/2^-}\, \bar{\Lambda} K^- \Xi^{*-} , \\
{\cal L}_{b(1/2^+)\to\Lambda K^-} &= g'_{1/2^+}\, \bar{\Lambda} i\gamma_5 \gamma^\mu \partial_\mu K^- b , \\
{\cal L}_{b(0^-)\to\Lambda \bar{\Xi}^+} &= g'_{0^-}\, \bar{\Lambda} \gamma_5 \bar{\Xi}^+ b ,
\end{align}
and the $g'_{J^P}$'s are the corresponding coupling constants. The propagators for the intermediate resonances are given by
\begin{align}
G^{\mu\nu}_{3/2}(q) &= i\,\frac{(q\mkern -9.5 mu /+m)\left[-g^{\mu\nu}+\frac{1}{3}\gamma^\mu\gamma^\nu+\frac{1}{3m^2}(q\mkern -9.5 mu /\gamma^\mu q^\nu + q^\mu\gamma^\nu q\mkern -9.5 mu /)\right]}{q^2-m^2+i m\Gamma}, \nonumber\\
G_{1/2}(q) &= i\,\frac{q\mkern -9.5 mu /+m}{q^2-m^2+i m\Gamma}, \nonumber\\
G_0(q) &= \frac{i}{q^2-m^2+i m\Gamma},
\end{align}
where $q$ is the four-momentum of the resonance, and $m$ and $\Gamma$ are the mass and total decay width of the intermediate $\Xi^*$ resonance. 
As in the experimental analysis~\cite{BESIII:2023mlv}, we also introduce form factors for the $\Xi^*$ resonances to account for their off-shell behavior.

In the calculation, we absorb the coupling constants $g_{J^P}$ and $g'_{J^P}$ for the relevant resonances, namely $\Xi(1690)$, $\Xi(1820)$, and the background, into the effective strength parameters $g$ listed in Table~I, i.e., $g_{\Xi(1690)}$, $g_{\Xi(1820)}$, and $g_{\rm bk}$. We emphasize that in the rescattering scenario, only the coupling constants appearing in the vertices of Eqs.~(11) and~(12) enter the definition of these $g$ parameters. The remaining dynamical information is encoded in the $T$-matrix, which is obtained separately from the coupled-channel interaction kernel.

\section{S2. Flavor structures of the Lagrangians under SU(3) symmetry}
In this section, we present the explicit flavor structures of the effective Lagrangians used in the main text. The interactions are constructed using SU(3) flavor symmetry as in Ref.~\cite{deSwart:1963pdg}. The expressions below list the flavor factors for vertices involving octet baryons ($BBV$), decuplet baryons ($B^*B^*V$), and transition couplings ($B^*BV$). The fields are denoted as follows: $N$, $\Xi$, $\Sigma$, $\Lambda$ are the octet baryons; $\Delta$, $\Xi^*$, $\Sigma^*$, $\Omega$ are the decuplet baryons; $\bm{\rho}$, $\omega$, $\phi$, $K^*$ are vector mesons. 

Starting with the octet–octet–vector interactions:
\begin{align}
{\cal L}_{BBV} &=
N^\dagger \bm{\tau} \cdot \bm{\rho} \, N \nonumber\\
&\quad + (2\alpha-1) \Xi^\dagger \bm{\tau} \cdot \bm{\rho} \, \Xi
+ (2\alpha-1)\Xi^\dagger \Xi \, \omega
+ \frac{4\sqrt{2}}{3}\alpha \,\Xi^\dagger \Xi \, \phi \nonumber\\
&\quad + 2\alpha \bigl[-i\bm{\Sigma}\times\bm{\Sigma}\cdot\bm{\rho}\bigr]
+ 2\alpha \,\Sigma^\dagger\Sigma\,\omega
+ \sqrt{2}(2\alpha-1)\,\Sigma^\dagger\Sigma\,\phi \nonumber\\
&\quad + \frac{2}{3}(5\alpha-2)\,\Lambda^\dagger\Lambda\,\omega
+ \frac{\sqrt{2}}{3}(2\alpha+1)\,\Lambda^\dagger\Lambda\,\phi \nonumber\\
&\quad + \sqrt{\frac{4}{3}}(1-\alpha)\bigl(\Lambda^\dagger\bm{\Sigma}\cdot\bm{\rho}
+ \bm{\Sigma}^\dagger\cdot\bm{\rho}\Lambda\bigr) \nonumber\\
&\quad - \sqrt{\frac{1}{3}}(1-4\alpha)\,g\,\bigl(\Xi^\dagger K_c\Lambda
+ \Lambda^\dagger K_c^\dagger \Xi\bigr) \nonumber\\
&\quad + \bigl(-\Xi^\dagger\bm{\Sigma}\cdot \bm{\tau} K_c
+ \text{h.c.}\bigr),
\end{align}
where $N=(p,n)^T$, $\Xi=(\Xi^0,\Xi^-)^T$, $\Sigma=(-\Sigma^+,\Sigma^0,\Sigma^-)$ and ${\bm \Sigma}=(\Sigma_1,\Sigma_2,\Sigma_3)$. $K_c=(\bar{K}^0,-K^-)^T$. Here ${\bm \rho}=(\rho_1,\rho_2,\rho_3)$ and $\rho^\pm=\frac{1}{\sqrt{2}}(\rho_1\mp i\rho_2)$. The notation ${\bm \Sigma}\cdot{\bm \rho}=\Sigma^+\rho^-+\Sigma^0\rho^0+\Sigma^-\rho^+$. $\bm{\tau}$ are the Pauli matrices acting in isospin space. Throughout we use the same $\alpha$ parameter ($\alpha=1$ in this work) consistently.
The physical $\eta$ and $\eta'$ mesons are mixtures of the SU(3) octet $\eta_8$ and singlet $\eta_1$, and similarly $\omega$ and $\phi$ are mixtures of $\omega_8$ and $\omega_1$:
\begin{align}
\begin{pmatrix} \eta \\ \eta' \end{pmatrix} &=
\begin{pmatrix} \cos\theta_P & -\sin\theta_P \\ \sin\theta_P & \cos\theta_P \end{pmatrix}
\begin{pmatrix} \eta_8 \\ \eta_1 \end{pmatrix}, \\
\begin{pmatrix} -\phi \\ \omega \end{pmatrix} &=
\begin{pmatrix} \cos\theta_V & -\sin\theta_V \\ \sin\theta_V & \cos\theta_V \end{pmatrix}
\begin{pmatrix} \omega_8 \\ \omega_1 \end{pmatrix}.
\end{align}
In this work we adopt $\eta = \eta_8$, $\eta' = \eta_1$ (i.e., $\theta_P = 0$). For vector mesons we assume ideal mixing with $\tan\theta_V = 1/\sqrt{2}$ ($\theta_V \approx 35.3^\circ$), which gives
\begin{align}
-\phi &= \sqrt{\frac{2}{3}} \omega_8 - \sqrt{\frac{1}{3}} \omega_1 = -s\bar{s},\\
\omega &= \sqrt{\frac{1}{3}} \omega_8 + \sqrt{\frac{2}{3}} \omega_1 = \sqrt{\frac{1}{2}} (u\bar{u}+d\bar{d}),
\end{align}
so that $\phi$ is a pure $s\bar{s}$ state and $\omega$ is a pure light-quark state. Note that our sign convention for the $\phi$ field differs from those in Refs.~\cite{Ronchen:2012eg,Oh:2004wp}; our $\phi$ is the negative of the field defined in those works. This choice ensures a consistent sign structure in the SU(3) matrix representation for the $VPP$ Lagrangian in Eq.~(1).

Next, the decuplet–decuplet–vector interactions:
\begin{align}
{\cal L}_{B^*B^*V} &=
\Delta^\dagger \bm{T} \cdot \bm{\pi} \Delta \nonumber\\
&\quad + \frac{1}{2} \Xi^{*\dagger} \bm{\tau}\cdot \bm{\rho} \Xi^* 
+ \Xi^{*\dagger} \Xi^* \, \omega
+ \sqrt{2}\,\Xi^{*\dagger} \Xi^* \, \phi \nonumber\\
&\quad + \bigl[-i\bm{\Sigma}^*\times\bm{\Sigma}^*\cdot\bm{\rho}\bigr]
+ \Sigma^{*\dagger}\Sigma^*\,\omega
+ \frac{1}{\sqrt{2}} \Sigma^{*\dagger}\Sigma^*\,\phi \nonumber\\
&\quad - \bigl( \Xi^{*\dagger} \bm{\Sigma}^* \cdot \bm{\tau} K_c
+ \text{h.c.} \bigr)
- \sqrt{\frac{3}{2}}\bigl( \Omega^\dagger K^\dagger \Xi^*
+ \Xi^{*\dagger} K \Omega \bigr) \nonumber\\
&\quad + \Omega^\dagger\Omega\,\omega
+ \sqrt{8}\,\Omega^\dagger\Omega\,\phi .
\end{align}
Here $\Delta = (\Delta^{++}, \Delta^+, \Delta^0, \Delta^-)^T$; the definitions of $\Xi^*$ and $\Sigma^*$ are analogous to those of $\Xi$ and $\Sigma$, respectively. The isospin transition matrix $\bm{T}$ for the $\Delta\pi$ vertex is given by the following components:
\begin{align}
T_{\Delta x} &= \sqrt{\frac{4}{15}} \begin{pmatrix}
0 & \frac{\sqrt{3}}{2} & 0 & 0 \\[2pt]
\frac{\sqrt{3}}{2} & 0 & 1 & 0 \\[2pt]
0 & 1 & 0 & \frac{\sqrt{3}}{2} \\[2pt]
0 & 0 & \frac{\sqrt{3}}{2} & 0
\end{pmatrix}, \nonumber\\
T_{\Delta y} &= -i \sqrt{\frac{4}{15}} \begin{pmatrix}
0 & \frac{\sqrt{3}}{2} & 0 & 0 \\[2pt]
-\frac{\sqrt{3}}{2} & 0 & 1 & 0 \\[2pt]
0 & -1 & 0 & \frac{\sqrt{3}}{2} \\[2pt]
0 & 0 & -\frac{\sqrt{3}}{2} & 0
\end{pmatrix}, \nonumber \\
T_{\Delta z} &= \sqrt{\frac{4}{15}} \begin{pmatrix}
\frac{3}{2} & 0 & 0 & 0 \\[2pt]
0 & \frac{1}{2} & 0 & 0 \\[2pt]
0 & 0 & -\frac{1}{2} & 0 \\[2pt]
0 & 0 & 0 & -\frac{3}{2}
\end{pmatrix}.
\end{align}
The definition of $\bm{T}$ in Ref.~\cite{Matsuyama:2006rp} differs by a factor of $\sqrt{15/4}$ from ours.

Now the transition couplings between decuplet and octet baryons:
\begin{align}
{\cal L}_{B^*BV} &=
\Delta^\dagger \bm{T}^\dagger \cdot \bm{\rho} \, N \nonumber\\
&\quad - \sqrt{6}\,\Xi^{*\dagger} \bm{\tau} \cdot \bm{\rho} \Xi
- \sqrt{6}\,\Xi^{*\dagger} \Xi \, \omega
+ \sqrt{3}\,\Xi^{*\dagger} \Xi \, \phi \nonumber\\
&\quad - \sqrt{6}\bigl[-i\bm{\Sigma}^*\times\bm{\Sigma}\cdot\bm{\rho}\bigr]
- \sqrt{6}\,\Sigma^{*\dagger}\Sigma\,\omega
+ \sqrt{3}\,\Sigma^{*\dagger}\Sigma\,\phi \nonumber\\
&\quad + \sqrt{6}\,\bigl(\Xi^{*\dagger}\bm{\Sigma}\cdot\bm{\tau}K_c\bigr)
+ \sqrt{2}\,\bigl(\Xi^{*\dagger}K_c\Lambda
+ \Sigma^{*\dagger}\cdot\bm{\rho}\Lambda\bigr) \nonumber\\
&\quad - \Omega^\dagger K^\dagger \Xi .
\end{align}
The isospin transition matrices for the $N\Delta$ vertex are:
\begin{align}
T_x &= \frac{1}{\sqrt{6}} \begin{pmatrix}
-\sqrt{3} & 0 & 1 & 0 \\
0 & -1 & 0 & \sqrt{3}
\end{pmatrix}, \\
T_y &= -\frac{i}{\sqrt{6}} \begin{pmatrix}
\sqrt{3} & 0 & 1 & 0 \\
0 & 1 & 0 & \sqrt{3}
\end{pmatrix}, \\
T_z &= \frac{1}{\sqrt{6}} \begin{pmatrix}
0 & 2 & 0 & 0 \\
0 & 0 & 2 & 0
\end{pmatrix}.
\end{align}

These expressions are used to compute the flavor factors $C_{ij}$ in the potential kernel.

\section{S3. Coefficients $C_{ij}$ and connection to chiral unitary approach}

First, we specify the flavor wave functions for the coupled channels used in the rescattering calculation. Since we consider the rescattering in the $K^-\Lambda$ channel, only states with total isospin $I=1/2$ and $I_z=-1/2$ are relevant. The wave functions for the various channels are constructed according to the following isospin Clebsch–Gordan decompositions:
\begin{align}
| \bar{K}\Lambda\rangle &= -|K^-\Lambda \rangle, \\
|\bar{K}\Sigma^{(*)} \rangle &= \sqrt{\frac{2}{3}}|\bar{K}^0\Sigma^{(*)-}\rangle + \sqrt{\frac{1}{3}}|K^-\Sigma^{(*)0}\rangle, \\
| \pi\Xi^{(*)}\rangle &= \sqrt{\frac{1}{3}}|\pi^0\Xi^{(*)-}\rangle - \sqrt{\frac{2}{3}}|\pi^-\Xi^{(*)0}\rangle, \\
|\eta\Xi^{(*)}\rangle &= |\eta\Xi^{(*)-}\rangle, \\
| K\Omega\rangle &=|K^0\Omega^- \rangle,
\end{align}
where we adopt the isospin phase conventions $K^- = -|1/2,-1/2\rangle$.

With above wave fucntion, the coefficients $C_{ij}$ for the transitions $MB\to MB$ and $MB^*\to MB^*$ can be obtianed and listed in Tables~\ref{tab:C1} and~\ref{tab:C2}, respectively; the coefficients for $MB\to MB^*$ can be obtained analogously. In the current work, we adopt $\alpha = 1$, and one can verify that the $C_{ij}$ coincide with those used in the chiral unitary approach after summing over all exchanges~\cite{Ramos:2002xh,Sarkar:2004jh}.

\renewcommand\tabcolsep{0.23cm}
\renewcommand{\arraystretch}{1.2}
\begin{table}[h!]
\caption{$C_{ij}$ for $MB\to MB$. Entries are given in the form ``$a, b, c$" for $\rho$, $\omega$ and $\phi$ exchange; a single number corresponds to $K^*$ exchange. \label{tab:C1}}  
\begin{tabular}{cccccc}
\toprule
& $\bar{K}\Sigma$ & $\bar{K}\Lambda$ & $\pi\Xi$ & $\eta\Xi$ \\ \hline
$\bar{K}\Sigma$  & $2\alpha,-\alpha, 2\alpha-1$ & $\alpha-1,0,0$ & $-\frac12$ & $\frac{3}{2}$ \\
$\bar{K}\Lambda$ & & $0,\frac{5\alpha-2}{3},\frac{-1-2\alpha}{3}$ & $\frac{1-4\alpha}{2}$ & $\frac{1-4\alpha}{2}$ \\
$\pi\Xi$        & & & $2(2\alpha-1),0,0$ & $0$  \\
$\eta\Xi$       & & & & $0$   \\
\bottomrule
\end{tabular}
\end{table}

\renewcommand\tabcolsep{0.46cm}
\renewcommand{\arraystretch}{1.2}
\begin{table}[h!]
\caption{$C_{ij}$ for $MB^*\to MB^*$. Entries are given in the same format as Table~\ref{tab:C1}. \label{tab:C2}}
\begin{tabular}{cccccc}
\toprule
& $\bar{K}\Sigma^*$ & $\pi\Xi^*$ & $\eta\Xi^*$ & $K\Omega$ \\ \hline
$\bar{K}\Sigma^*$  & $2,-1,1$ & $-1$ & $3$ & $0$ \\
$\pi\Xi^*$        & & $2,0,0$ & $0$ & $-\frac{3}{\sqrt{2}}$ \\
$\eta\Xi^*$       & & & $0,0,0$ & $\frac{3}{\sqrt{2}}$ \\
$K\Omega$         & & & & $0,-1,4$ \\ \bottomrule
\end{tabular}
\end{table}

The current form of the potential can be reduced to that obtained in the chiral unitary approach~\cite{Ramos:2002xh,Sarkar:2004jh}. The reduction is performed by assuming $q^2 \ll m_V^2$ and neglecting the $q^\mu q^\nu$ term in the propagator, which leads to a contact interaction. The resulting potential simplifies to
\begin{align}
V_{MB\to MB} &= -\frac{C_{ij}}{4f^2}\, \bar{u}\,\gamma^\nu (k+k')_\nu u,\\
V_{MB^*\to MB^*} &= -\frac{C_{ij}}{4f^2}\, \bar{u}^\rho\,\gamma^\nu (k+k')_\nu u_\rho,
\end{align}
where we have used the relation $g = m_V/(2f)$ with $f \approx 93$~MeV the pseudoscalar decay constant~\cite{Aceti:2014uea}. Under this approximation, after choosing $\alpha=1$, the present potential reduces completely to that employed in the chiral unitary approach~\cite{Ramos:2002xh,Sarkar:2004jh}.

\section{S4. qBSE approach for rescattering scenario}

The rescattering amplitude ${\cal T}$, depicted in Figs.~1 (a), is constructed by substituting the potential kernel into the Bethe-Salpeter equation. Through the application of the quasipotential approximation together with a partial wave decomposition, the original four-dimensional integral equation defined in Minkowski space can be transformed into a one-dimensional integral equation that applies separately to each spin-parity $J^P$ channel,
\begin{align}
&i{\cal T}^{J^P}_{\lambda_1,\lambda_2;\lambda'_1,\lambda'_2}({\rm p}',{\rm p}) \nonumber\\
&= i{\cal V}^{J^P}_{\lambda_1,\lambda_2;\lambda'_1,\lambda'_2}({\rm p}',{\rm p}) + \frac12 \sum_{\lambda''_1,\lambda''_2} \int \frac{{\rm p}''^2 d{\rm p}''}{(2\pi)^3} \nonumber\\
&\quad \cdot i{\cal V}^{J^P}_{\lambda_1,\lambda_2;\lambda''_1,\lambda''_2}({\rm p}',{\rm p}'') \, G_0({\rm p}'') \, i{\cal T}^{J^P}_{\lambda''_1,\lambda''_2;\lambda'_1,\lambda'_2}({\rm p}'',{\rm p}), \label{Eq: TJP}
\end{align}
where the symbols $\lambda_{1,2}$, $\lambda'_{1,2}$, and $\lambda''_{1,2}$ represent the helicities of the incoming, outgoing, and intermediate mesons and baryons, respectively.

The propagator $G_0({\rm p}'')$ is obtained by reducing its original four-dimensional version using the quasipotential approximation, and its explicit expression reads,
\begin{align}
G_0({\rm p}'') &= \frac{\delta^+(p''^2_h - m_h^2)}{p''^2_l - m_l^2} \nonumber\\
&= \frac{\delta^+(p''^0_h - E_h({\rm p}''))}{2E_h({\rm p}'')\left[(W - E_h({\rm p}''))^2 - E_l^2({\rm p}'')\right]},
\end{align}
Here $p''_h$ and $p''_l$ denote the four-momenta of the heavier and the lighter particle in the intermediate state, respectively, while $W$ is the total center-of-mass energy. In the present work, we employ the spectator approximation, where the heavier particle (indicated by $h$) is taken on its mass shell, i.e., $p''^0_h = E_h({\rm p}'') = \sqrt{m_h^2 + {\rm p}''^2}$. As a result, the energy of the lighter particle $l$ is fixed by $p''^0_l = W - E_h({\rm p}'')$. Throughout this paper, the magnitude of the three-momentum in the center-of-mass frame is written as ${\rm p} = |{\bm p}|$.

The partial-wave projected potential ${\cal V}^{J^P}_{\lambda_1,\lambda_2;\lambda'_1,\lambda'_2}({\rm p}',{\rm p})$ is derived from the original interaction potential ${\cal V}_{\lambda_1,\lambda_2;\lambda'_1,\lambda'_2}({\bm p}',{\bm p})$ via the following projection formula,
\begin{align}
{\cal V}_{\lambda_1,\lambda_2;\lambda'_1,\lambda'_2}^{J^P}({\rm p}',{\rm p}) &= 2\pi \int d\cos\theta \Big[ d^{J}_{\lambda'_{21},\lambda_{21}}(\theta) \, {\cal V}_{\lambda_1,\lambda_2;\lambda'_1,\lambda'_2}({\bm p}',{\bm p}) \nonumber\\
&\quad + \eta \, d^{J}_{-\lambda'_{21},\lambda_{21}}(\theta) \, {\cal V}_{\lambda_1,\lambda_2;-\lambda'_1,-\lambda'_2}({\bm p}',{\bm p}) \Big],\label{Eq:VJP}
\end{align}
In this expression, $d^{J}_{\lambda\lambda'}(\theta)$ are the Wigner $d$-functions that characterize the rotation between the initial and final helicity states. These functions depend on the helicity difference $\lambda_{21} = \lambda_2 - \lambda_1$, and $\theta$ is the angle between the initial and final three-momenta. The factor $\eta$ is defined as $\eta = P P_1 P_2 (-1)^{J - J_1 - J_2}$, where $P$ stands for the intrinsic parity of the two-particle system, $P_1$ and $P_2$ are the parities of the two constituents, and $J$, $J_1$, $J_2$ are the total spin and the individual spins, respectively. To facilitate the partial wave expansion, we choose the initial three-momentum as ${\bm p} = (0,0,{\rm p})$ and the final three-momentum as ${\bm p}' = ({\rm p}'\sin\theta, 0, {\rm p}'\cos\theta)$. In addition, an exponential form factor is incorporated into the propagator to regularize the high-momentum behavior as
\[
G_0({\rm p}'') \rightarrow G_0({\rm p}'') \, e^{-2(p''^2_l - m_l^2)^2 / \Lambda^4},
\]
where $\Lambda$ is the cutoff parameter that governs the strength of the regularization.

The total decay amplitude corresponding to Fig.~1(a) is obtained by incorporating the rescattering amplitude ${\cal T}$ into the decay process:
\begin{align}
i{\cal M}_{\lambda_1,\lambda_2;\lambda_3;\lambda}(p_1,p_2,p_3)&=i\int \frac{d^4p'_2}{(2\pi)^4} \, {\cal T}_{\lambda_1,\lambda_2}(p_1,p_2;p'_1,p'_2) \nonumber\\
& \cdot G(p'_2)\, {\cal A}_{\lambda_3;\lambda}(p'_1,p'_2,p_3),
\end{align}
where $\lambda$ and $\lambda_i$ denote the helicities of the initial $\psi(3686)$ and the final $K^-$, $\Lambda$, and $\Xi^+$, respectively. The momenta $p_1/p'_1$ and $p_2/p'_2$ correspond to the final $K^-$, $\Lambda$ (intermediate particles), respectively, while $p_3$ is the momentum of the $\Xi^+$. $G(p'_2)$ is the propagator of the intermediate particle.

Since the rescattering amplitude is obtained in a partial-wave-decomposed form, a partial wave decomposition of the decay amplitude is also performed. Including parity, the decay amplitude can be rewritten as
\begin{align}
&i{\cal M}_{\lambda_1,\lambda_2,\lambda_3;\lambda}(p_1,p_2,p_3)
\nonumber\\
&=\sum_{J^PM}\frac{2J+1}{4\pi}D^{J*}_{M\lambda_{21}}(\Omega_2)\int \frac{{\rm p}'^{2}d{\rm p}'}{(2\pi)^3} \nonumber\\
&\cdot\sum_{\lambda'_1\lambda'_2} i{\cal T}^{J^P}_{\lambda_1,\lambda_2;\lambda'_1\lambda'_2}({\rm p}'_j,s_{12})  \ G_0({\rm p}') i{\cal A}^{J^PM}_{\lambda'_1\lambda'_2;\lambda_3;\lambda}({\rm p}'_j,\Omega_3,s_{12}).
\end{align}
Here, ${\cal T}^{J^P}$ is the partial-wave-decomposed rescattering amplitude in (\ref{Eq: TJP}). The angles $\Omega_2$ and $\Omega_3$ represent the solid angles of the final $\Lambda$ and $\Xi^+$, respectively, while $s_{12}$ is the invariant mass of the $K^-\Lambda$ system. The Wigner $D$-function $D^{J*}_{M\lambda_{21}}(\Omega_2)$ depends on the Euler angles $\Omega_2$.

\section{S5. Invariant mass spectra in the Breit-Wigner scenario}

Here we present the three invariant mass spectra obtained in the Breit-Wigner scenario in Fig.~\ref{fig:BW}.
.

\begin{figure*}[t]
\centering
\includegraphics[bb=5 0 1050 700,clip,scale=0.48]{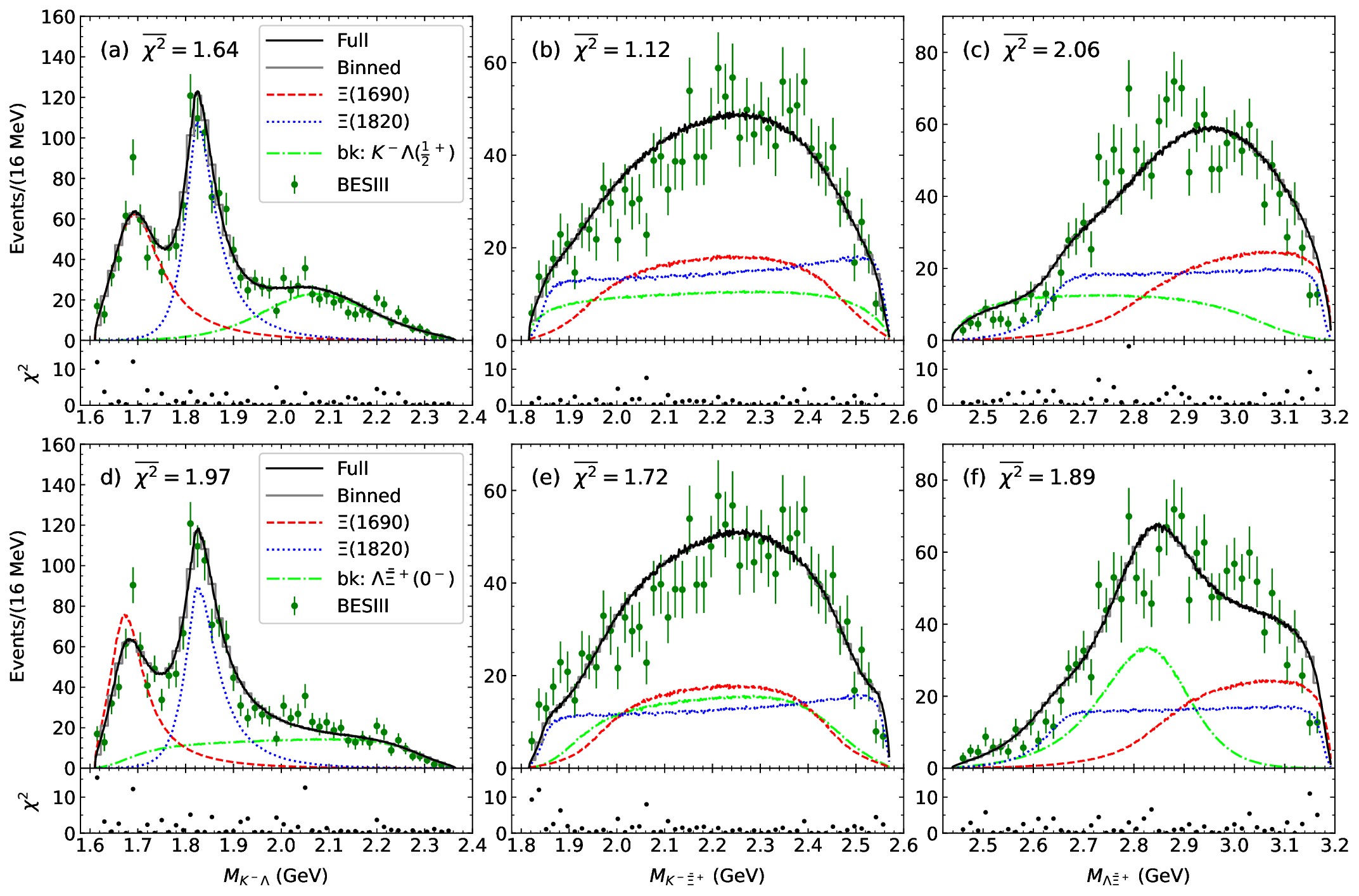}
\caption{Invariant mass spectra in the Breit-Wigner scenario. The curves correspond to the full model (solid black), $\Xi(1690)$ (red dashed), $\Xi(1820)$ (blue dotted), and background (green dash-dotted). The results binned into the data points are also shown as a gray histogram. The data are taken from the BESIII Collaboration~\cite{BESIII:2023mlv}. The scatter points in the lower part of each panel represent the individual $\chi^2$ contributions of the corresponding data points. Panels (a,b,c) correspond to the fit with a background in $K^-\Lambda$ ($1/2^-$), and (d,e,f) to the fit with a background in $\Lambda\bar{\Xi}^+$ ($0^-$).}
\label{fig:BW}
\end{figure*}

As in the experimental analysis, the data can be well reproduced with two $\Xi^*$ states and a background in $K^-\Lambda$ with $1/2^-$. The average $\chi^2$ values for the $K^-\Lambda$, $K^-\bar{\Xi}^+$, and $\Lambda\bar{\Xi}^+$ invariant mass spectra are 1.64, 1.12, and 2.06, respectively. These values are close to those obtained in the experimental analysis (1.46, 1.10, and 2.16)~\cite{BESIII:2023mlv}. As shown in Fig.~\ref{fig:BW}(a), the peaks around 1700 and 1800~MeV in the $K^-\Lambda$ spectrum clearly originate from the two Breit-Wigner resonances, while the background mainly contributes to the higher energy region. For the $K^-\bar{\Xi}^+$ spectrum in Fig.~\ref{fig:BW}(b), the three contributions interfere significantly, affecting a broad energy range, and their shapes differ due to different spin-parities and central masses. In the $\Lambda\bar{\Xi}^+$ spectra shown in Fig.~\ref{fig:BW}(c), the contributions from $\Xi(1690)$ and $\Xi(1820)$ appear at higher energies, with the lower-mass $\Xi(1820)$ contributing to a lower energy region, whereas the background, having a larger central mass, contributes to an even lower energy region.

One finds that the largest $\chi^2$ comes from the $\Lambda\bar{\Xi}^+$ invariant mass spectrum. Therefore, we replace the background in $K^-\Lambda$ with $1/2^-$ by a background in $\Lambda\bar{\Xi}^+$ with $0^-$. The $\chi^2$ for the $\Lambda\bar{\Xi}^+$ spectrum improves slightly from 2.06 to 1.89. However, the average $\chi^2$ values for the $K^-\Lambda$ and $K^-\bar{\Xi}^+$ spectra increase from 1.64 and 1.12 to 1.97 and 1.72, respectively. Moreover, the background contribution in $K^-\Lambda$ becomes broader. The peak around 1700~MeV in Fig.~\ref{fig:BW}(d) is almost the same as that in Fig.~\ref{fig:BW}(a), but the fitted mass of the $\Xi(1690)$ is 1667.3~MeV, which is much smaller than the 1682.9~MeV obtained with the background in $K^-\Lambda$ with $1/2^-$. This suggests a large interference. These results are also consistent with the experimental analysis~\cite{BESIII:2023mlv}, where the background in $K^-\Lambda$ with $1/2^-$ is the best choice.

An important observation is that the experimental peak around 1700~MeV is very sharp. Neither of the fits shown in Fig.~\ref{fig:BW}(a) and (d) can reproduce such a sharp peak; in fact, even the original experimental analysis using Breit-Wigner resonances failed to describe it~\cite{BESIII:2023mlv}. This provides strong evidence that the peak does not originate from a conventional Breit-Wigner resonance, but rather points to a non-resonant or dynamically generated nature.

\section{S6. Cutoff dependence of the results  near $\Xi(1690)$}
\label{sec:cutoff}

\begin{figure}[h!]
\centering
\includegraphics[bb=5 0 1050 700,clip,scale=0.45]{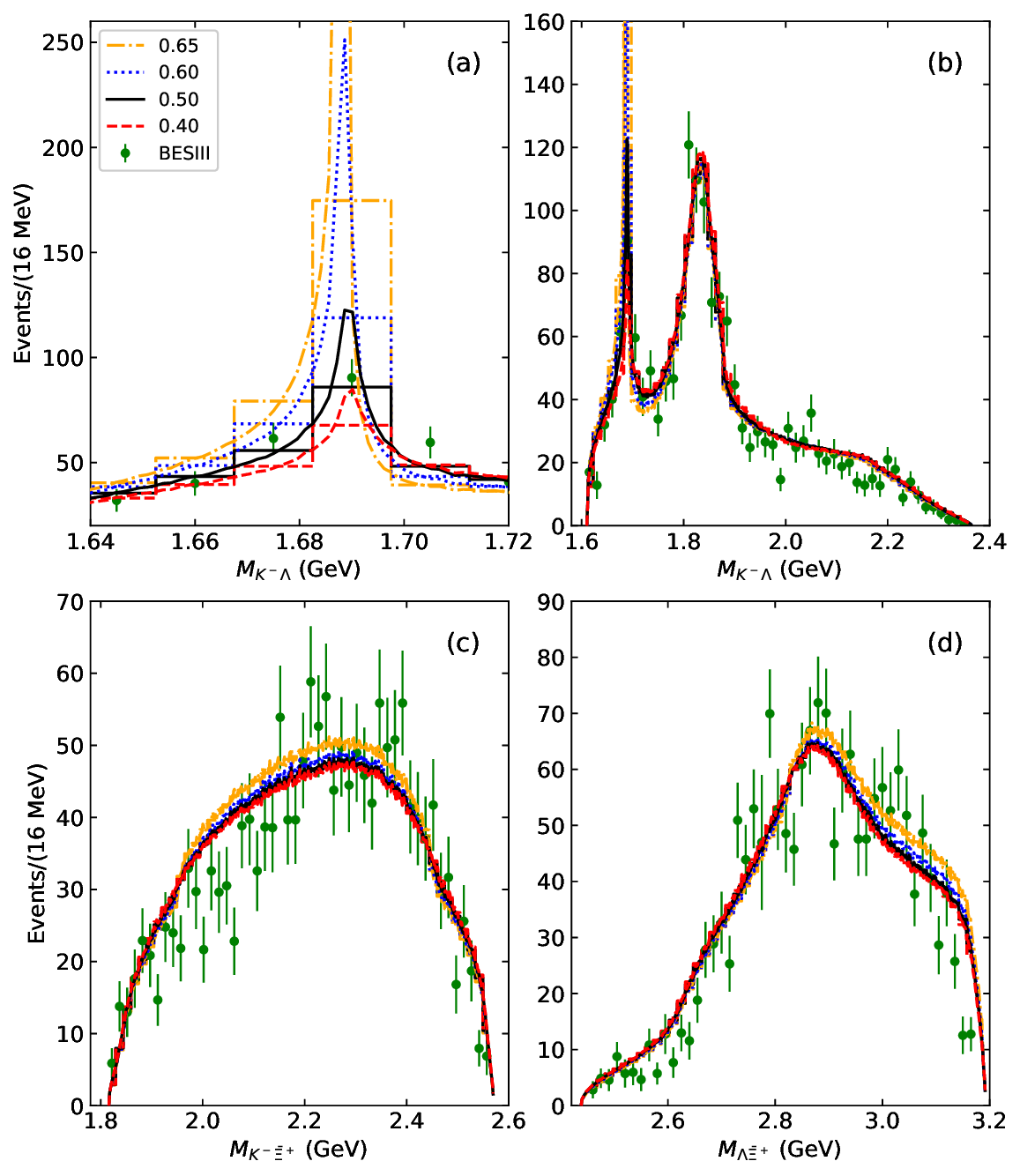}
\caption{Invariant mass spectra of the $K^-\Lambda$ system in the $\Lambda\bar{\Xi}^+$ ($0^-$) background region near the $\Xi(1690)$, obtained with different cutoff parameters $\Lambda$ for the $\bar{K}\Sigma$ channel. The curves correspond to $\Lambda=0.50$ (orange dash-dotted), $0.55$ (blue dotted), $0.60$ (green dashed), and $0.65$ (red solid) GeV. The binned data points from the BESIII Collaboration~\cite{BESIII:2023mlv} are shown as histograms for comparison.}
\label{fig:cutoff}
\end{figure}

To examine the sensitivity of our results to the cutoff parameter, we vary $\Lambda$ for the $\bar{K}\Sigma$ channel and compute the corresponding $K^-\Lambda$ invariant mass spectra. We focus on the scenario with the $\Lambda\bar{\Xi}^+$ ($0^-$) background, which provides the best description of the experimental data. The cutoff is varied around the best-fit value $\Lambda = 0.50$ GeV, namely $\Lambda = 0.40$, $0.50$, $0.55$, $0.60$, and $0.65$ GeV. The resulting spectra are shown in Fig.~\ref{fig:cutoff}.

With the best-fit value $\Lambda = 0.50$ GeV, a virtual state is dynamically generated at about 1.7 MeV below the $\bar{K}\Sigma$ threshold. The resulting peak in the $K^-\Lambda$ spectrum is slightly higher than the data points, but after binning, the theoretical values fall within the uncertainties of the experimental data, indicating a reasonable agreement.

When the cutoff is decreased to $\Lambda = 0.40$ GeV, the interaction becomes weaker, and the virtual state moves further away from the threshold, to about 5.3 MeV below. Since the virtual state resides on the second Riemann sheet, its influence on the physical $K^-\Lambda$ spectrum is reduced. Consequently, the peak lies below the central values of the data, and the binned results fall outside the experimental uncertainties.

If we increase the cutoff to $\Lambda = 0.60$ GeV, the pole moves very close to the threshold, at about 0.004 MeV below, behaving effectively as a bound state. In this case, the peak becomes significantly higher than the data, and the binned results exceed the uncertainty bands. For even larger cutoffs, such as $\Lambda = 0.65$ GeV, the pole becomes a bound state at about 0.1 MeV below threshold, and the resulting peaks are further enhanced. The discrepancy with the data becomes even more pronounced, with clear deviations also appearing in the $K^-\bar{\Xi}$ and $\Lambda\bar{\Xi}$ spectra.

These observations clearly demonstrate that the peak height increases rapidly with increasing cutoff. The data strongly favor the virtual-state scenario, since any bound-state interpretation would produce a peak much higher than observed. We therefore conclude that our results are not only stable against reasonable variations of the cutoff, but also provide strong evidence that the $\Xi(1690)$ is a virtual state generated by the $\bar{K}\Sigma$ interaction.

Here we also provide a physical interpretation for the choice of the cutoff parameters and their differences. First, the cutoffs for different resonances should reflect the underlying interaction scales; the closer the natures of the systems, the closer the cutoffs should be. In our results, the $\Xi(1690)$ arises mainly from the $\bar{K}\Sigma$ interaction (a pseudoscalar meson with an octet baryon), while the $\Xi(1820)$ originates from the $K\Sigma^*$ and $\pi\Xi^*$ interactions (a pseudoscalar meson with a decuplet baryon). Given these distinct underlying dynamics, it is natural that the effective cutoff scales differ. Second, we adopt a simplified prescription in which the cutoffs for all other channels are kept fixed, while only the cutoff for the dominant channel is varied to fit the data. As shown above, a change of $\Lambda$ by 0.25~GeV for the $\bar{K}\Sigma$ channel shifts the pole position by only a few MeV near the threshold, but produces a dramatic effect in the invariant mass spectrum. This demonstrates that the invariant mass spectra provide a sensitive constraint on the cutoff parameters, and the extracted values are meaningful despite their apparent difference.


\begin{thebibliography}{99}


\bibitem{ParticleDataGroup:2024cfk}
S.~Navas \textit{et al.} [Particle Data Group],
Phys. Rev. D \textbf{110}, 030001 (2024)

\bibitem{Dalitz:1960du}
R.~H.~Dalitz and S.~F.~Tuan,
Annals Phys. \textbf{10}, 307-351 (1960)

\bibitem{Hyodo:2011ur}
T.~Hyodo and D.~Jido,
Prog. Part. Nucl. Phys. \textbf{67}, 55-98 (2012)

\bibitem{Oset:1997it}
E.~Oset and A.~Ramos,
Nucl. Phys. A \textbf{635}, 99-120 (1998)

\bibitem{Hall:2014uca}
J.~M.~M.~Hall, W.~Kamleh, D.~B.~Leinweber, B.~J.~Menadue, B.~J.~Owen, A.~W.~Thomas and R.~D.~Young,
Phys. Rev. Lett. \textbf{114}, no.13, 132002 (2015)

\bibitem{LEPS:2003wug}
T.~Nakano \textit{et al.} [LEPS],
Phys. Rev. Lett. \textbf{91}, 012002 (2003)

\bibitem{Wu:2010jy}
J.~J.~Wu, R.~Molina, E.~Oset and B.~S.~Zou,
Phys. Rev. Lett. \textbf{105}, 232001 (2010)

\bibitem{Yang:2011wz}
Z.~C.~Yang, Z.~F.~Sun, J.~He, X.~Liu and S.~L.~Zhu,
Chin. Phys. C \textbf{36}, 6-13 (2012)


\bibitem{LHCb:2015yax}
R.~Aaij \textit{et al.} [LHCb],
Phys. Rev. Lett. \textbf{115}, 072001 (2015)

\bibitem{LHCb:2019kea}
R.~Aaij \textit{et al.} [LHCb],
Phys. Rev. Lett. \textbf{122}, no.22, 222001 (2019)

\bibitem{Wu:2009tu}
J.~J.~Wu, S.~Dulat and B.~S.~Zou,
Phys. Rev. D \textbf{80}, 017503 (2009)

\bibitem{Wang:2024jyk}
E.~Wang, L.~S.~Geng, J.~J.~Wu, J.~J.~Xie and B.~S.~Zou,
Chin. Phys. Lett. \textbf{41}, no.10, 101401 (2024)

\bibitem{He:2025vij}
J.~He,
Phys. Rev. C \textbf{112}, no.1, 015205 (2025)

\bibitem{Amsterdam-CERN-Nijmegen-Oxford:1976ezm}
J.~B.~Gay \textit{et al.} [Amsterdam-CERN-Nijmegen-Oxford],
Phys. Lett. B \textbf{62}, 477-480 (1976)


\bibitem{Teodoro:1978bu}
D.~Teodoro, J.~Diaz, C.~Dionisi, J.~B.~Gay, R.~J.~Hemingway, M.~J.~Losty, M.~Mazzucato, R.~Blokzijl, G.~G.~G.~Massaro and H.~Voorthuis, \textit{et al.}
Phys. Lett. B \textbf{77}, 451-453 (1978)

\bibitem{Biagi:1986vs}
S.~F.~Biagi, M.~Bourquin, R.~M.~Brown, H.~J.~Burckhart, P.~Extermann, M.~Gailloud, C.~N.~P.~Gee, W.~M.~Gibson, P.~Jacot-Guillarmod and J.~Perrier, \textit{et al.}
Z. Phys. C \textbf{34}, 175 (1987)

\bibitem{BaBar:2008myc}
B.~Aubert \textit{et al.} [BaBar],
Phys. Rev. D \textbf{78}, 034008 (2008)

\bibitem{Belle:2018lws}
M.~Sumihama \textit{et al.} [Belle],
Phys. Rev. Lett. \textbf{122}, no.7, 072501 (2019)

\bibitem{BESIII:2015dvj}
M.~Ablikim \textit{et al.} [BESIII],
Phys. Rev. D \textbf{91}, no.9, 092006 (2015)



\bibitem{BESIII:2023mlv}
M.~Ablikim \textit{et al.} [BESIII],
Phys. Rev. D \textbf{109}, no.7, 072008 (2024)

\bibitem{Capstick:1986ter}
S.~Capstick and N.~Isgur,
Phys. Rev. D \textbf{34}, no.9, 2809-2835 (1986)

\bibitem{Glozman:1995fu}
L.~Y.~Glozman and D.~O.~Riska,
Phys. Rept. \textbf{268}, 263-303 (1996)



\bibitem{Ramos:2002xh}
A.~Ramos, E.~Oset and C.~Bennhold,
Phys. Rev. Lett. \textbf{89}, 252001 (2002)

\bibitem{Sarkar:2004jh}
S.~Sarkar, E.~Oset and M.~J.~Vicente Vacas,
Nucl. Phys. A \textbf{750}, 294-323 (2005)
[erratum: Nucl. Phys. A \textbf{780}, 90-90 (2006)]

\bibitem{Khemchandani:2016ftn}
K.~P.~Khemchandani, A.~Mart{\'\i}nez Torres, A.~Hosaka, H.~Nagahiro, F.~S.~Navarra and M.~Nielsen,
Phys. Rev. D \textbf{97}, no.3, 034005 (2018)

\bibitem{Sekihara:2015qqa}
T.~Sekihara,
PTEP \textbf{2015}, no.9, 091D01 (2015)

\bibitem{Feijoo:2023wua}
A.~Feijoo, V.~Valcarce Cadenas and V.~K.~Magas,
Phys. Lett. B \textbf{841}, 137927 (2023)
[erratum: Phys. Lett. B \textbf{853}, 138660 (2024)]

\bibitem{Kolomeitsev:2003kt}
E.~E.~Kolomeitsev and M.~F.~M.~Lutz,
Phys. Lett. B \textbf{585}, 243-252 (2004)


\bibitem{Molina:2023uko}
R.~Molina, W.~H.~Liang, C.~W.~Xiao, Z.~F.~Sun and E.~Oset,
Phys. Lett. B \textbf{856}, 138872 (2024)

\bibitem{Liang:2024fsv}
W.~H.~Liang, R.~Molina and E.~Oset,
Phys. Rev. D \textbf{110}, no.3, 036005 (2024)


\bibitem{Oller:2000fj}
J.~A.~Oller and U.~G.~Meissner,
Phys. Lett. B \textbf{500}, 263-272 (2001)

\bibitem{Jido:2003cb}
D.~Jido, J.~A.~Oller, E.~Oset, A.~Ramos and U.~G.~Meissner,
Nucl. Phys. A \textbf{725}, 181-200 (2003)


\bibitem{Albaladejo:2016lbb}
M.~Albaladejo, P.~Fernandez-Soler, F.~K.~Guo and J.~Nieves,
Phys. Lett. B \textbf{767}, 465-469 (2017)


\bibitem{Guo:2017jvc}
F.~K.~Guo, C.~Hanhart, U.~G.~Mei{\ss}ner, Q.~Wang, Q.~Zhao and B.~S.~Zou,
Rev. Mod. Phys. \textbf{90}, no.1, 015004 (2018)
[erratum: Rev. Mod. Phys. \textbf{94}, no.2, 029901 (2022)]

\bibitem{Roca:2005nm}
L.~Roca, E.~Oset and J.~Singh,
Phys. Rev. D \textbf{72}, 014002 (2005)

\bibitem{Geng:2006yb}
L.~S.~Geng, E.~Oset, L.~Roca and J.~A.~Oller,
Phys. Rev. D \textbf{75}, 014017 (2007)

\bibitem{He:2015cca}
J.~He and P.~L.~Lu,
Int. J. Mod. Phys. E \textbf{24}, no.11, 1550088 (2015)



\bibitem{Bando:1984ej}
M.~Bando, T.~Kugo, S.~Uehara, K.~Yamawaki and T.~Yanagida,
Phys. Rev. Lett. \textbf{54}, 1215 (1985)

\bibitem{Birse:1996hd}
M.~C.~Birse,
Z. Phys. A \textbf{355}, 231-246 (1996)

\bibitem{deSwart:1963pdg}
J.~J.~de Swart,
Rev. Mod. Phys. \textbf{35}, 916-939 (1963)
[erratum: Rev. Mod. Phys. \textbf{37}, no.2, 326-326 (1965)]



\bibitem{Ronchen:2012eg}
D.~Ronchen, M.~Doring, F.~Huang, H.~Haberzettl, J.~Haidenbauer, C.~Hanhart, S.~Krewald, U.~G.~Meissner and K.~Nakayama,
Eur. Phys. J. A \textbf{49}, 44 (2013)

\bibitem{Janssen:1996kx}
G.~Janssen, K.~Holinde and J.~Speth,
Phys. Rev. C \textbf{54}, 2218-2234 (1996)

\bibitem{Oh:2004wp}
Y.~Oh, K.~Nakayama and T.~S.~H.~Lee,
Phys. Rept. \textbf{423}, 49-89 (2006)

\bibitem{Matsuyama:2006rp}
A.~Matsuyama, T.~Sato and T.~S.~H.~Lee,
Phys. Rept. \textbf{439}, 193-253 (2007)

\bibitem{Lenske:2016ymj}
H.~Lenske, M.~Dhar and T.~Gaitanos,
[arXiv:1602.08917 [nucl-th]].


\bibitem{He:2019rva}
J.~He and D.~Y.~Chen,
Eur. Phys. J. C \textbf{79}, no.11, 887 (2019)


\bibitem{He:2014nya}
J.~He,
Phys. Rev. D \textbf{90}, no.7, 076008 (2014)


\bibitem{He:2015mja}
J.~He,
Phys. Rev. D \textbf{92}, no.3, 034004 (2015)

\bibitem{He:2017lhy}
J.~He and D.~Y.~Chen,
Eur. Phys. J. C \textbf{78}, no.2, 94 (2018)

\bibitem{He:2015yva}
J.~He,
Phys. Rev. C \textbf{91}, no.1, 018201 (2015)

\bibitem{He:2015cea}
J.~He,
Phys. Lett. B \textbf{753}, 547-551 (2016)



\bibitem{Duan:2024ygq}
M.~Y.~Duan, J.~Song, W.~H.~Liang and E.~Oset,
Eur. Phys. J. C \textbf{84}, no.9, 947 (2024)

\bibitem{James:1968gu}
F.~James,
``Monte-Carlo phase space,''
CERN-68-15 (1968)

\bibitem{code}
https://github.com/junhe1979/DalitzPlot.jl

M.~Ablikim \textit{et al.} [BESIII],
Phys. Rev. D \textbf{109}, no.7, 072008 (2024)

\bibitem{Ramos:2002xh}
A.~Ramos, E.~Oset and C.~Bennhold,
Phys. Rev. Lett. \textbf{89}, 252001 (2002)

\bibitem{Sarkar:2004jh}
S.~Sarkar, E.~Oset and M.~J.~Vicente Vacas,
Nucl. Phys. A \textbf{750}, 294-323 (2005)
[erratum: Nucl. Phys. A \textbf{780}, 90-90 (2006)]


\bibitem{Aceti:2014uea}
F.~Aceti, M.~Bayar, E.~Oset, A.~Martinez Torres, K.~P.~Khemchandani, J.~M.~Dias, F.~S.~Navarra and M.~Nielsen,
Phys. Rev. D \textbf{90}, no.1, 016003 (2014)


\bibitem{deSwart:1963pdg}
J.~J.~de Swart,
Rev. Mod. Phys. \textbf{35}, 916-939 (1963)
[erratum: Rev. Mod. Phys. \textbf{37}, no.2, 326-326 (1965)]



\bibitem{Ronchen:2012eg}
D.~Ronchen, M.~Doring, F.~Huang, H.~Haberzettl, J.~Haidenbauer, C.~Hanhart, S.~Krewald, U.~G.~Meissner and K.~Nakayama,
Eur. Phys. J. A \textbf{49}, 44 (2013)


\bibitem{Oh:2004wp}
Y.~Oh, K.~Nakayama and T.~S.~H.~Lee,
Phys. Rept. \textbf{423}, 49-89 (2006)

\bibitem{Matsuyama:2006rp}
A.~Matsuyama, T.~Sato and T.~S.~H.~Lee,
Phys. Rept. \textbf{439}, 193-253 (2007)



\end{thebibliography}
\end{document}